
%
%
%
%
%
%
%
\input harvmac

\def\rone{{r_1}}
\def\rtwo{{r_2}}
\def\zst{Z_{\rm string}}
\def\CF{{\cal F}}
\def\pl{Phys. Lett. }
\def\ceff{c_{\rm eff}}
\def\ao{A_{\rm open}}
\Title{PUPT--1277}
{{\vbox {\centerline{Some Properties of (Non) Critical Strings
}}}}

\bigskip
\centerline{D. Kutasov\footnote{*}
{\rm Lectures
at the ICTP Spring School on String Theory and Quantum Gravity, Trieste,
April 1991.}}
\bigskip\centerline
{\it Joseph Henry Laboratories,}
\centerline{\it Princeton University,}
\centerline{\it Princeton, NJ 08544.}

\vskip .2in

\noindent
We review some recent developments in string theory, emphasizing
the importance of vacuum instabilities, their relation to the density
of states, and the role of space-time fermions in non-critical string
theory. We also discuss the classical dynamics
of two dimensional string theory.

\Date{9/91}
%

\newsec{INTRODUCTION.}

In the last few years many people have
been studying simple models of string theory,
which correspond to string propagation in two dimensional space-time
(and simple generalizations thereof). Of course, for most applications
\ref\GSW{M. Green, J. Schwarz and E. Witten, {\it Superstring theory},
Cambridge Univ. Press (1987) and references therein.},
\ref\GAUGE{A. Polyakov, {\it Gauge Fields and Strings}, Harwood
Academic Publishers (1987).} one needs to consider much more complicated
models, however many important issues in string theory are still not
understood, and the hope is that the two dimensional theory will serve
as a useful toy model, in which some of these issues may be addressed.
Due to the low dimension of space-time, the number of degrees of freedom
in the theory is vastly smaller than in the twenty six (or ten) dimensional
case. The physical on shell states include one field theoretic degree
of freedom, the ``tachyon'' center of mass of the string (which is actually
massless in two dimensions), and a discrete infinite set of massive states --
the remnants of the tower of oscillator states in $D>2$ (where the number
of field theoretic degrees of freedom with mass $\leq m_0$ diverges
exponentially with $m_0$).

Due to the absense of transverse excitations, one might expect the dynamics
to be simpler in two dimensions. Indeed, following ideas of
\ref\MATMOD{
E. Brezin, C. Itzykson, G. Parisi and J.B. Zuber, Comm. Math. Phys.
{\bf 59} (1978) 35 ; D. Bessis, Comm. Math. Phys. {\bf 69} (1979) 147}
\ref\SURF{V. Kazakov, Phys. Lett. {\bf 150B} (1985) 282;
F. David, Nucl. Phys. {\bf B257} (1985) 45; J. Ambjorn, B. Durhuus and
J. Frohlich, Nucl. Phys. {\bf B257} (1985) 433; V. Kazakov, I. Kostov
and A. Migdal, Phys. Lett. {\bf 157B} (1985) 295.}
it has been shown
\ref\ORIG{D. Gross and A. Migdal, Phys. Rev. Lett. {\bf 64}
(1990) 127; M. Douglas and S. Shenker, Nucl. Phys. {\bf B335} (1990) 635;
E. Brezin and V. Kazakov, Phys. Lett. {\bf B236} (1990) 144.}
\ref\CONE{
E. Brezin, V. Kazakov, and Al. Zamolodchikov, Nucl. Phys. {\bf B338}
(1990) 673; D. Gross and N. Miljkovi\'c, Phys. Lett {\bf 238B} (1990) 217;
P. Ginsparg and J. Zinn-Justin, Phys. Lett {\bf 240B} (1990) 333;
G. Parisi, Phys. Lett. {\bf 238B} (1990) 209.}
using large $N$ matrix model techniques,
that many properties of two dimensional (2d) strings are exactly
calculable to all orders in the topological (genus) expansion. The
simplicity of these models is closely related to an underlying
free fermion structure \MATMOD\ --
\ref\BDSS{T. Banks, M. Douglas, N. Seiberg and S. Shenker,
Phys. Lett. {\bf 238B} (1990) 279.}.

There are two main potential domains of
application of the results of \ORIG, \CONE.
One is two dimensional (world sheet)
quantum gravity. The implications of matrix models
for quantum gravity have been recently discussed in
\ref\NATI{N. Seiberg, Rutgers preprint RU-90-29 (1990).},
\ref\MSS{G. Moore, N. Seiberg and M. Staudacher, Rutgers
preprint RU-91-11 (1991).},
\ref\GREGNATI{G. Moore and N. Seiberg, Rutgers preprint RU-91-29 (1991).}.
The second application, which will be of main interest to us here, is to
critical (unified) string theory. The two are of course closely connected,
and we will mention some aspects of this relation as we go along. The approach
we will take is to try and understand the matrix model results in the
continuum path integral formalism \GSW,
and use this understanding to shed light on the structure of the
theory, in particular its space-time dynamics. This program is far from
complete; we will review what has been achieved and discuss some of the
open problems.

We will take a broader point of view, presenting 2d strings
in the context of higher dimensional string theory, which is ultimately
what we're interested in. This will also help to try and identify
features of general validity, and those that are special to two dimensions.
With that in mind, we will also investigate some string models
in two dimensions which are more difficult to treat (or in some cases
even formulate) using matrix models.

In the continuum formulation of 2d quantum gravity
\ref\POL{A. Polyakov, Phys. Lett. {\bf B103} (1981) 207, 211.}
\ref\KPZ{
V. Knizhnik,
A. Polyakov and A. Zamolodchikov, Mod. Phys. Lett {\bf A3} (1988) 819.},
we are given a general conformal field theory (CFT) with action
${\cal S}_M(g)$ (on a Riemann surface with metric $g_{ab}$), central
charge $c_M$, and we have to integrate over all metrics $g$
on manifolds of fixed topology (genus). In the conformal gauge
$g_{ab}=e^\phi {\hat g}_{ab}$, the dynamical metric $g$ is represented
by the Liouville mode $\phi$; the action is:
\eqn\a{ {\cal S} ={\cal S}_L({\hat g}) + {\cal S}_M({\hat g}) }
where
\eqn\SL{ {\cal S}_L({\hat g})={1 \over {2 \pi}} \int \sqrt{\hat g}[
{\hat g}^{ab} \partial_a \phi \partial_b \phi -{Q \over 4}{\hat R} \phi
+2 \mu e^{\alpha_+ \phi}] }
$Q$ and $\alpha_+$ are parameters which are fixed by gauge invariance
\ref\DDK{F. David, Mod. Phys. Lett. {\bf A3} (1988) 1651;
J. Distler and H. Kawai, Nucl. Phys. {\bf B321} (1989) 509.}. It is
very useful to think about the Liouville mode as a target space cordinate,
so that \a\ is a particular background of the critical string system
in two dimensions
\ref\IND{S. Das, S. Naik and S. Wadia, Mod. Phys. Lett. {\bf A4} (1989)
1033; J. Polchinski, Nucl. Phys. {\bf B324} (1989) 123;
T. Banks and J. Lykken, Nucl. Phys. {\bf B331} (1990) 173.}.
Therefore, we have to set the total central charge of matter ($c_M$) and
Liouville ($c_L=1+3Q^2$) to 26:
\eqn\QA{Q=\sqrt{25-c_M\over3}}
Spinless matter primary operators of dimension $\Delta(=\bar\Delta)$ $V_\Delta$
acquire Liouville dependence,
\eqn\Td{T_\Delta(\phi)=V_\Delta\exp(\beta_\Delta\phi)}
with
\eqn\mass{\Delta-{1\over2}\beta(\beta+Q)=1}
In particular, putting $\Delta=0$ in \mass\ fixes $\alpha_+$ in \SL.
As usual in critical string theory, \mass\ is the mass shell condition
(or the linearized equation of motion for small excitations). The
vertex operators $T_\Delta$ are related to the wave functions of the
corresponding states $\Psi_\Delta$
by $T_\Delta(\phi)=g_{\rm st}(\phi)\Psi_\Delta(\phi)$,
where unlike the ``usual'' cases \GSW, the string coupling $g_{\rm st}$
is {\it not} constant in general \SL; rather we have $g_{\rm st}=
g_0\exp(-{Q\over2}\phi)$. Therefore,
\eqn\ps{\Psi_\Delta=V_\Delta e^{(\beta+{Q\over2})\phi}}
and $E=\beta+{Q\over2}$ is identified as the Liouville momentum of the
state described by $\Psi$.
The form \ps\ is actually only an asymptotic approximation of the exact
wave function. It is valid in the region $e^{\alpha_+\phi}<<1$
($\phi\rightarrow\infty$ in our conventions). The exact wave function
in two dimensional string theory is known \MSS\ to satisfy the Wheeler
de Witt equation,
\eqn\WdW{\left[-{\partial^2\over\partial\phi^2}+\mu\exp(\alpha_+\phi)
+\nu^2\right]\Psi_\Delta(\phi)=0}
where $\nu^2=2\Delta-{c_M-1\over12}$. Eq. \WdW\ was derived
\MSS\ in the matrix model approach. In the continuum formalism,
\WdW\ holds in a minisuperspace approximation; it was not yet derived
in the full theory.
It is very natural to interpret $\phi$ as
Euclidean time. Eq.
\mass\ takes then the form:
\eqn\dmass{E^2=m^2;\;\;m^2=2\Delta-{c_M-1\over12}}
The wave function $\Psi_\Delta$ with energy $E$
describes one quantum mechanical degree of freedom
in space-time. If, as in the minimal models of
\ref\BPZ{A. Belavin, A. Polyakov and A. Zamolodchikov, Nucl. Phys.
{\bf B241}, 333 (1984).},
the matter system contains
a finite number of spinless primaries $V_\Delta$, the corresponding
string theory reduces to quantum mechanics of a finite number of degrees
of freedom. In general, of course, the number of matter
primaries is enormous, and the number of degrees of freedom
is much larger than in field theory. If there are some
non compact dimensions $X_i$, such that $V_\Delta(X)=e^{ik\cdot X}\tilde
V_h$, $\Delta={1\over2} k^2+h$, we can rewrite \dmass\ as
$$E^2=k^2+m^2,\;\;
m^2=2h-{c_M-1\over12}$$ and $V$ describes one field theoretic degree of
freedom in the appropriate dimension.

{}From \dmass\ we see that operators with $\Delta<{c_M-1\over24}$
in the matter CFT correspond in space-time to on shell tachyons, which
lead (as in field theory) to IR instabilities.
In field theory, these IR divergences
imply that we are expanding around the wrong vacuum, and should
move to a new, stable one. We will see that in string theory, as it is
currently understood, the situation is surprisingly different.

{}From the point of view of 2d gravity, tachyons correspond to normalizable
wave functions \ps, while states with positive $m^2$ are described
by wave functions which are exponentially supported at $\phi\rightarrow
\pm\infty$ (and are of course non normalizable). The world sheet
interpretation of this \NATI\ is that normalizable (tachyonic)
wave functions are
supported mainly on surfaces with finite size holes {\it in the dynamical
metric} $g_{ab}$; they describe macroscopic states, and if we perturb the
action \a\ by the corresponding operators, the dynamical surface
is destabilized by the multitude of holes that are created. Non normalizable
wave functions with $E>0$ are supported on very small surfaces in the
dynamical metric, and therefore describe small disturbances of the
surface (microscopic states). The corresponding operators can be (and were)
studied in the matrix model approach. Wave functions with $E<0$
describe very large disturbances of the surface, and it is not known how
to treat them in the matrix model (although there are some plausible
suggestions).

Thus, there is a nice correspondence between IR instabilities in space
time (tachyons) and on the world sheet (macroscopic states which create
holes in the surface). The role of states with $E<0$ is not well
understood at this time, and is in fact one of the important remaining
problems, which is related to many properties of the theory, as we will
see below. Note that by \dmass\ the familiar relation between tachyons
and relevant matter operators is only true for $c_M=25$. In general, for
$c_M<25$ relevant operators with ${c_M-1\over24}<\Delta<1$ do not create
instabilities; for $c_M>25$ we have the unintuitive situation that
irrelevant matter operators with $1<\Delta<{c_M-1\over24}$ are tachyonic.
The region $c_M>25$ has been less studied than $c_M<25$, and we will not
discuss it further here.

For tachyonic operators, $\beta$ in \Td\ is complex \dmass. In the original
work of \KPZ, \DDK\ (see also
\ref\CT{T. Curtright and C. Thorn, Phys. Rev. Lett. {\bf 48} (1982) 1309;
E. Braaten, T. Curtright and C. Thorn, Phys. Lett. {\bf 118B} (1982) 115;
Ann. Phys. {\bf 147} (1983) 365.}
\ref\GN{J.L Gervais and A. Neveu, Nucl. Phys. {\bf B209} (1982) 125;
{\bf B224} (1983) 329; {\bf B238} (1984) 125.}) it was noted
that the identity operator ($V_\Delta=1$ in \Td) becomes tachyonic
when $c_M>1$. Since the gravitational coupling constant is proportional
to $1\over26-c_M$ \POL, it has been suggested that at $c_M=1$ a certain
transition between ``weak gravity'' $c_M<1$ and ``strong gravity''
$c_M>1$ takes place. Taken naively, this point of view was not
completely satisfactory, since it is easy to construct matter theories
where $c_M<1$, but there exist tachyonic operators in the spectrum;
it is not likely that the identity operator plays a special role in this
context. The issue was later clarified in \NATI,
\ref\NADA{D. Kutasov and N. Seiberg, Nucl. Phys. {\bf B358} (1991) 600.};
the current understanding is that the crucial quantity is not the
gravitational coupling constant, but rather the {\it density of states}
of the theory. Gravity has ${1\over2}n(n-3)$ degrees of freedom
in $n$ dimensions; for $n=2$ this is $-1$. If the matter theory
has more than ``one degree of freedom'', in a sense which we will make precise
below, the full string theory contains tachyons. Therefore, the
invariant way to describe the famous ``$c=1$ barrier'' of \KPZ, is that
{\it 2d gravity coupled to matter with more than one field theoretic degree of
freedom is unstable}.
The relation between the existence of tachyonic excitations around
a certain vacuum, and the number of states in it implies that if it
is possible at all to turn on expectation values of the various fields
and move from a tachyonic vacuum to a stable one in string theory,
the new stable vacuum is always ``trivial''
(at least in bosonic string theory).

Hence, the density of states is of major importance in string theory, as are
tachyonic instabilities. Section 2 of these notes is devoted to a precise
definition of the density of states and its relation to the presence of
tachyons and stability. We also discuss the implications of this
relation for string dynamics.

Since the only stable vacua of bosonic string theory (where by bosonic
we mean vacua with only bosonic space-time excitations; e.g. the fermionic
string falls into this cathegory as well) are those which are essentially
two dimensional field theories in space-time, we are led in section 3
to consider theories with space-time fermions
\ref\KUS{D. Kutasov and N. Seiberg, Phys. Lett. {\bf B251} (1990) 67.}.
We construct explicitly a large set of theories with a number of degrees
of freedom varying between that of a two dimensional field theory and that
of the ten dimensional superstring, all of which are stable solutions
of the equations of motion of string theory. This is possible since space
time fermions contribute a negative amount to the ``number of degrees
of freedom'', and a theory with many bosons and many fermions can still
have a vanishingly small total ``density of states''. This number is required
to be small for stability.
The theories thus obtained have many of the favorable properties
of critical superstring models, but exhibit some puzzling features
as well
(such as continuous breaking of SUSY),
mostly related to their time dependence.

In section 4 we describe the known facts about classical dynamics in two
dimensional string theory. One of the issues of interest is the exact
form of the classical equations of motion of the string. The usual way
one obtains those in string theory is by studying
scattering amplitudes of on shell fields. Calculating the correlation
functions of $T_\Delta$ \Td\ involves solving the interacting
Liouville theory \SL. Despite many efforts \CT, \GN, \NATI,
\ref\JP{J. Polchinski, Texas preprints UTTG-19-90; UTTG-39-90 (1990).},
this is still an open problem. What saves the day is that there exists a class
of amplitudes which contain most of the information, and are simply
calculable. To understand this, it is useful to focus on the dynamics of
the zero mode of the Liouville field $\phi$ (the minisuperspace
approximation in 2d gravity). The main source of the complications
in Liouville theory is the ``cosmological'' interaction in \SL.
{}From the space-time point of view, this corresponds
to a (zero $X$ momentum) tachyon condensate, while on the world sheet
this is a potential for $\phi$. The reason why it is needed is to
keep the Liouville field away from the region $\phi\rightarrow-\infty$
where the string coupling $g_{\rm st}\rightarrow\infty$. Many amplitudes
diverge when $\mu\rightarrow0$ in \SL. By KPZ scaling \KPZ, the amplitudes
$\langle T_{\Delta_1}...T_{\Delta_n}\rangle$ are proportional to
$\mu^s$, where
\eqn\Spar{\sum_{i=1}^N\beta_i+\alpha_+s=-Q}
In general, all amplitudes are non zero -- there is no Liouville
momentum conservation \Spar. Instead, the $\phi$ zero mode integral,
which can be explicitly performed, yields
\ref\GTW{ A. Gupta, S. Trivedi and M. Wise, Nucl. Phys. {\bf B340}
(1990) 475.}:
\eqn\zeromode{\langle T_{\Delta_1}...T_{\Delta_n}\rangle=
\left({\mu\over\pi}\right)^s\Gamma(-s)
\langle T_{\Delta_1}...T_{\Delta_n}\left[\int\exp(\alpha_+\phi)\right]^s
\rangle_{\mu=0}}
where the correlator on the r.h.s. is understood to exclude the Liouville
zero mode.
One can see from \zeromode\ that
the amplitudes with $s=0$ are special. The $\phi$ zero mode integral
is given for them by:
$$\int_{-\infty}^\infty d\phi_0\exp(-\mu e^{\alpha_+\phi_0})$$
This diverges from $\phi\rightarrow\infty$; imposing a UV cutoff
$\phi_0\leq\phi_{UV}$, we find
$\log\mu$ multiplied by a free field amplitude for the
Liouville field (i.e. Liouville is treated as a Feigin-Fuchs field,
and the cosmological interaction is absent). Of course, we still
have to multiply by the matter contribution and integrate over moduli,
but this reduces the calculation to well defined integrals; similar
representations for generic Liouville amplitudes
\zeromode\ are not known.

Why does this simplification of the amplitudes with $s=0$ occur?
A major clue is provided by the observation that these amplitudes
are insensitive to the particular form of the ``wall'' that keeps
the Liouville field away from $\phi=-\infty$. For example, replacing
$e^{\alpha_+\phi}$ by a general $T_\Delta$ would do the job, and while
generic amplitudes depend strongly on the choice of the condensate $T_\Delta$,
the $s=0$ ones do not (apart from trivial overall factors). The existence
of the Liouville wall is of course a boundary effect, hence the $s=0$
amplitudes correspond to {\it bulk scattering}, which is clearly
independent of the form of the boundary at strong coupling.
The reason for the simplicity is that in the bulk of the
$\phi$ volume, all the complications of Liouville theory are
irrelevant, and the Liouville field is free. The overall factor of
$\log{1\over\mu}$ which appears in all bulk amplitudes is interpreted
as the volume of the Liouville coordinate, as appropriate for bulk effects.

A similar simplification occurs in \zeromode\ for all $s\in Z_+$. The
power of $\int\exp(\alpha_+\phi)$ on the r.h.s.
becomes then a positive integer;
the divergence of $\mu^s\Gamma(-s)$ is interpreted again as\foot{
This can be obtained by starting with the fixed area amplitude, using
$\mu^s\Gamma(-s)=\int_0^\infty dA A^{-s-1} e^{-\mu A}$. The divergence
at $s\in Z_+$ is interpreted then as a small area divergence in the integral
over areas, which can be regulated by a small area (UV) cutoff, as before,
yielding the stated result.}: $\mu^s\Gamma(-s)\simeq\mu^s\log\mu$.
The space-time interpretation is as before: now it involves bulk
interaction of the scattering particles $T_{\Delta_1}...T_{\Delta_n}$, with
the zero momentum tachyons that form the Liouville wall.

The bulk amplitudes provide us with the required probe of the dynamics
of $D$ dimensional string theories; they can be defined
and evaluated
for all string vacua (all matter CFT's $M$),
while general Liouville amplitudes \zeromode\ are expected to be much more
subtle for $D>2$. The bulk amplitudes should also be sufficient to
deduce the equations of motion of the theory, and compare the dynamics
around different vacua. In principle one can construct from them
a space-time Lagrangian which gives those amplitudes, and then
incorporate boundary effects to deduce all the amplitudes of the theory
\ref\PO{A. Polyakov, Mod. Phys. Lett. {\bf A6} (1991) 635.}.

In view of the above discussion, section 4 is devoted
to the bulk tree level
S -- matrix in two dimensional string theory.
Surprisingly, the moduli integrals can be explicitly evaluated in this
case, essentially due to the same simplification of the dynamics of the
theory, as that leading to the free fermion structure of the corresponding
matrix models. In the continuum formalism, the simplicity is due to
a partial decoupling of a certain infinite set of discrete string states.
We review the results and mention some problems still left in the continuum
approach to 2d strings.
We also compare the situation in 2d closed string theory to that found
in open 2d string theory. Section 5 contains some concluding remarks.
\bigskip

\newsec{DENSITY OF STATES AND TACHYONS IN STRING THEORY.}

In this section we will show that the density of states
(or number of degrees of freedom) of the system plays a central role
in two dimensional quantum gravity and string theory. In particular,
we will show that matter with `too many' degrees of freedom can not
be consistently coupled to gravity (under mild assumptions) due to the
appearance of tachyons. We believe that
this restriction on the number of states in quantum mechanical
generally covariant theories is more general.

We start by defining the density of states. The most general situation
we are interested in is an arbitrary vacuum of critical string theory,
i.e. an arbitrary conformal field theory with $c=26$ (or
$\hat c=10$). As we
saw in section 1, gravity coupled to a matter CFT with any $c_M$
is a particular example, with the missing central charge $26-c_M$
supplied by the Liouville CFT. We'll start by considering this special
case, and then generalize. To count states it is convenient to evaluate
the torus partition sum:
\eqn\Zt{Z(\tau)={\rm Tr}q^{L_0-{c\over24}}{\bar q}^{\bar L_0-{c\over24}}}
where $q=e^{2\pi i\tau}$, and $\tau=\tau_1+i\tau_2$ is the complex modulus
of the torus.

The string partition sum \Zt\ factorizes into a product of three
contributions, from ghosts (for which a factor of $(-)^G$, $G$ -- ghost
number, must be inserted in \Zt), Liouville and the ``matter'' CFT.
The ghosts contribute to $Z(\tau)$ a factor of
$|\eta(\tau)|^4$. The Liouville contribution is that
of a free scalar field \GTW\ $\log\mu\left(\sqrt{\tau_2}|\eta(\tau)|^2
\right)^{-1}$; the cosmological term in \SL\ enters trivially,
for the same reasons as in section 1 -- the one loop free energy
is a bulk amplitude; hence it is proportional to the volume $\log\mu$,
with the coefficient given by a free field amplitude.
Finally, the matter is described by a partition sum $Z_{\rm CFT}(\tau)$
(defined as in \Zt).

The total partition sum is (after multiplying by a factor of $\tau_2$ for later
convenience)
\eqn\strings{\zst =\sqrt{\tau_2} |\eta|^2 Z_{\rm CFT}  }
Modular invariance of $\zst(\tau)$ is a fundamental principle
in string theory. We will assume it throughout our discussion. In ordinary
CFT, the partition sum \Zt\ can be used to count states in the theory.
One considers the behavior of the partition sum ($Z_{\rm CFT}$ or
$\zst$), when $\tau_1=0, \tau_2=\beta\rightarrow 0$, where
$Z(\beta)={\rm Tr}\;e^{-\beta E}$ has the form:
\eqn\zdef{Z(\beta)\sim \beta^r
\exp({\pi\ceff \over 6\beta})
}
Clearly, $\ceff$ is a measure of the number of degrees of freedom
of the theory, since in the limit
$\beta\rightarrow0$ all states contribute 1 to $Z(\beta)$
\Zt. By modular invariance $Z(\beta)=Z({1\over\beta})$,
and \Zt, $\ceff$
is related to the lowest lying states in the spectrum:
\eqn\ceffec{\eqalign{
\ceff  &= c-24 \Delta_m \cr
\Delta_m &={1\over 2} {\rm min} (\Delta +\bar \Delta)  \cr}
}
In unitary CFT $\Delta_m=0$, and $\ceff=c$. So the central charge
measures the number of degrees of freedom
\ref\cthem{A. B. Zamolodchikov JETP Lett. {\bf 43} (1986) 730; J. Cardy
and A. Ludwig, Nucl. Phys. {\bf B285[FS19]} (1987) 687;
A. Cappelli, D. Friedan and J. Latorre, Rutgers preprint
RU-90-43 (1990).}. In non unitary theories, $\Delta_m$ is generically
negative and $\ceff>c$. In Liouville theory, $\ceff=1$ ($\Delta_m>0$). Thus
$c(=1+3Q^2)$ is not a good measure of the density of states, which
is that of a free scalar field for all $Q$.

In string theory, $\ceff$ is not a good measure of the density of
states since
not all states contributing to \zdef\ are physical --
we have to impose the constraint $L_0=\bar L_0$. This is implemented
by integrating over $\tau_1$; we define:
\eqn\go{G(\tau_2) = \int _{-{1\over 2}}^{1\over 2} d\tau_1\zst }
By modular invariance we know that $\zst(\tau_1+1,\tau_2)=\zst(\tau_1,
\tau_2)$, hence only states of integer spin contribute to $\zst$
$(\Delta-\bar\Delta\in Z)$. Transforming to $G(\tau_2)$ \go\ elliminates
contributions of all states except those with $\Delta=\bar\Delta$.
The function $G(\tau_2)$ has a simple space-time interpretation. It is
related to the one loop free energy of the particle excitations of the
string: ${\rm Tr}(-)^F\log(p^2+m^2)$ ($F$ is the space-time fermion number).
In a proper time representation, the one loop free energy $\Omega$ is given
by:
\eqn\sch{\Omega={\rm Tr} (-)^F\log(p^2+m^2)=
\int_0^\infty {ds\over s} \sum_n (-1)^{F_n} \int d^D p e^{-s(p^2+m_n^2)}}
where $p$ is the momentum in the non compact directions and the sum
over $n$ runs over the space-time excitations. The function $G(\tau_2)$
of \go\ is related to the integrand in \sch:
\eqn\gs{G({s\over2\pi})=s\sum_n(-)^{F_n}\int d^Dp e^{-s(p^2+m_n^2)}}
One consequence of this is that all states should contribute positive amounts
to $G(\tau_2)$, unless there are space-time fermions in the spectrum.
{}From the definition of $G(\tau_2)$ \strings, \go, this is far from clear.
While the CFT partition sum as well as that of Liouville satisfy this
property, being traces over Hilbert spaces with positive weights, ghost
oscillators flip the sign of the contributions to \Zt\ (due to the
factor of $(-)^G$ in the definition of the
ghost partition sum). Therefore, even in
bosonic string theory there may appear physical states at non trivial
ghost numbers, which contribute a negative amount to $G(\tau_2)$ \go.
Such states indeed do appear at discrete values of the momenta $p$.
They seem to decouple from the dynamics due to the ghost numbers, and their
role is not completely clear. In more than two dimensions the statistical
weight of these states is low and they can be ignored (except perhaps
as generators of symmetries). In two dimensional space-time they seem to be
closely related to the symmetry structure of the theory
\ref\witten{E. Witten, IAS preprint IASSNS-HEP-91/51 (1991).}. In any event,
the field theoretic degrees of freedom, over which the sum in \gs\ runs, occur
at zero ghost number and contribute with positive sign to the partition sum.

A state with $\Delta=\bar\Delta$ in the CFT contributes to $G(\tau_2)$,
an amount $e^{-4\pi\tau_2(\Delta-{c_M-1\over24})}$.
If $\Delta<{c_M-1\over24}$,
the Schwinger integral \sch\ developes an IR divergence (from
${s\over2\pi}=\tau_2\rightarrow\infty$).
This IR divergence is due to the tachyon
instability. Hence, the behavior of $G(\tau_2)$ as $\tau_2\rightarrow
\infty$ probes the existence of tachyons in the spectrum. On the other
hand, its behavior as $\tau_2\rightarrow0$ is a measure of the density
of states
of the theory, since in that limit all states contribute
1 to $G$. Now, the theories we are describing here all have an infinite
number of states, therefore generically $G$ diverges as $\tau_2\rightarrow
0$; however we can estimate the density of states by measuring how fast
$G$ diverges in that limit.

On general grounds we know that as $\tau_2\rightarrow0$,
$G(\tau_2)\simeq \tau_2^x e^{y\over\tau_2}$. Therefore it is natural
to define:
\eqn\cstring{c_{\rm string}-1= \lim_{\tau_2 \rightarrow 0 }
{6\tau_2\over\pi}\log G(\tau_2)}
as a measure of the number of degrees of freedom. The $-1$ on the
left hand side of \cstring\ reflects the contribution of the gravity
sector to the density of states mentioned above. If one takes, e.g.,
the matter CFT to be one scalar field or a minimal model \BPZ,
it is easy to see that $c_{\rm string}=1$. Thus the minimal models
and two dimensional string theory have the same density of states.
In general, $c_{\rm string}>1$ for theories without space-time fermions.
We will see soon that such theories always contain tachyons.

As explained above, tachyons cause an IR divergence in $\Omega$,
thus finiteness of $\Omega$ is a good measure of the existence of tachyons
(in fermionic string theories this is strictly true only for theories
without tachyonic space-time fermions, e.g. theories with unitary
matter). Of course, the field theoretic expression for $\Omega$ \sch\
has in addition to the IR tachyon divergence which we are interested
in, a UV divergence (from $s=0$). This is usually delt with
by a proper time cutoff: $\int_0^\infty ds\rightarrow \int_{\Lambda^{-2}}
^\infty ds$. String theory removes this divergence by a different
mechanism \GSW. One notes that by \go, \sch, \gs,
$\Omega=\int{d^2\tau\over\tau_2^2} \zst(\tau)$, where the integral runs
over the half infinite strip $\tau_2\geq0, \;|\tau_1|\leq{1\over2}$.
This may diverge due to modular invariance of $\zst$ since the strip
contains an infinite number of copies of a fundamental domain of the
modular group. If this is the only source of divergence,
one can cure it by restricting the integral
to a fundamental domain $\CF$ of the modular group, $|\tau|\geq1$,
thus avoiding the ``UV region'' $\tau_2\rightarrow0$. But this
``stringy regularization'' actually teaches us something very
interesting about the theory. Suppose there are no tachyons in the
spectrum. Then the integral $\int_{\CF} {d^2\tau\over\tau_2^2}\zst$
is finite, and the only possible source of divegrence
of the integral over the strip is the volume of the modular group.
We can try to estimate this divergence by using the ``field theoretic''
cutoff: integrate $\int {d^2\tau\over\tau_2^2}$ over the cutoff
strip: $|\tau_1|\leq{1\over2}$, $\tau_2>{1\over\Lambda^2}$. The
integral of $\zst$ over the cutoff strip
should diverge at the same rate as the (regularized) volume
of the modular group, which is given by $\int_{\Lambda^{-2}}^\infty
{d\tau_2\over\tau_2^2}\int_{-{1\over2}}^{1\over2}d\tau_1 1=\Lambda^2$.
Hence we learn that from the space-time point of view, tachyon free
string theories have a free energy $\Omega(\Lambda)$, which diverges
(at most) as $\Omega\simeq \Lambda^2$ as we remove the cutoff.
But in space-time the rate of divergence of $\Omega$ measures
the density of states of the theory. The behavior we find
in tachyon free string theory indicates a very small number of states.
Even a theory with one bosonic field in $D$ dimensions
would have $\Omega(\Lambda)\simeq\Lambda^D$. In string theory we generically
have an infinite number of space-time fields, so the behavior we
find
is even more unusual. We would expect $\Lambda^a e^{b\Lambda}$ in general,
due to the Hagedorn-like growth of the number of states with mass.
By the above arguments, such theories are always tachyonic. The only
theories that can be tachyon free are those that exhibit the general
features of 2D field theories!

Pushing these ideas one step further, we can derive a more quantitative
correspondence between $\Omega$ and the number of states. If the divergence
of \sch\ is indeed due to the volume of the modular group, we should
have a relation of the form:
\eqn\Reg{\lim_{\Lambda\rightarrow\infty}{\int_{\Lambda^{-2}}^\infty
{d\tau_2\over\tau_2^2}\int_{-{1\over2}}^{1\over2}d\tau_1 \zst\over
\int_{\Lambda^{-2}}^\infty
{d\tau_2\over\tau_2^2}\int_{-{1\over2}}^{1\over2}d\tau_1 1}=
{\int_{\CF}{d^2\tau\over\tau_2^2}\zst\over\int_{\CF}{d^2\tau\over\tau_2^2}
1}}
Equation \Reg\ is the statement that the two regularizations of
$\Omega$ using the field theoritic proper time cutoff, and the
stringy modular
invariant cutoff, are equivalent. This is in fact only true under
certain assumptions, which we will soon state, but assuming it is true,
we derive by evaluating the l.h.s. the following relation:
\eqn\pro{\int_\CF {d^2\tau \over \tau_2^2} \zst= \lim_{\Lambda\rightarrow
\infty}{\pi \over 3 \Lambda^2 }\int_{\Lambda^{-2}}^\infty {d\tau_2 \over
\tau_2^2 }G(\tau_2) =\lim_{\Lambda\rightarrow \infty}
{\pi \over 3} G(\Lambda^{-2})}
This relation states that the (regularized) one loop free energy is equal
(up to a constant) to
the number of states of the theory $G(0)$. Before going on
to prove \pro, we would like to make several comments on its
significance and implications.

Eq. \pro\ is a statement about modular invariant functions (obeying certain
conditions to be stated below). We have presented it for the
case of a CFT coupled to Liouville, however the discussion applies
to all string vacua which are described by modular invariant partition
sums. Important generalizations include fermionic strings, superstrings,
and heterotic strings, as well as arbitrary CFT's with $c=26$.

The l.h.s. of \pro\ diverges iff there are tachyons in the spectrum. Hence,
if there are no tachyons, the r.h.s. must also be finite. Looking back
at \gs, we see that this implies that the theory has the number of states
of a two dimensional field theory with a finite number of fields. If there
are no space-time fermions, this means that the theory contains far
fewer states than generic string theories \GSW. The only bosonic
string theories without tachyons are two dimensional (e.g. the $c=1$
or minimal models coupled to gravity, and the coset model of
\ref\WBH{E. Witten, Phys. Rev. {\bf D44} (1991) 314.}). This means that
the role of tachyons in string theory is more fundamental
than in field theory. Unlike there, it doesn't seem likely that tachyons
can be gotten rid of by shifting to a nearby vacuum.
For that to happen, one of two scenarios must occur: either
space-time fermions must somehow dynamically appear, or the new
stable vacuum must effectively be a two dimensional string theory.
Starting e.g. from the 26 dimensional bosonic string, neither
possibility seems likely.
In general, it seems unlikely that turning on expectation
values of fields can change $c_{\rm string}$, which should be invariant
under small perturbations.
The fate of string vacua containing
tachyons (or 2D quantum gravity systems with ``too much'' matter)
remains unclear.

Fermionic string theories with a non chiral GSO projection contain also
only bosonic excitations in space-time. Hence the situation there is
completely analogous to the bosonic case -- tachyon free theories have
very few states (the number of states is again as in two dimensional field
theory). Theories with many states contain tachyons. It is well
known that non trivial theories exist in this case, but those involve
a {\it chiral} GSO projection (e.g. the ten dimensional superstring \GSW).
It is important to note that such theories are not necessarily space-time
supersymmetric (one example of a non supersymmetric theory is the $O(16)
\times O(16)$ theory \GSW). However, it is known that they always
contain space-time fermions. Our discussion uncovers a surprising
feature of these theories: even though space-time SUSY is absent
in general, tachyon free superstrings are {\it asymptotically
supersymmetric}. In other words, while the number of
bosons and fermions is not the same energy level by energy level,
at high enough energy, the total numbers of bosons and fermions
up to that energy are the same to fantastic precision. More
precisely, in this case it is natural to write $G(s)$ \gs\ as
$G(s)=G_B(s)-G_F(s)$, where $G_B$ and $G_F$ measure the contributions
of bosons and fermions to the free energy \gs, and the relative minus
sign is due to spin statistics; then in general
we have $G_B(s), \;G_F(s)\simeq s^a e^{b\over s}$, while
$G_B(s)-G_F(s)\rightarrow {\rm const}$
as $s\rightarrow0$
(again, by using \pro\ and finiteness
of its l.h.s.). Hence bosons and fermions almost
cancel in any tachyon free string theory. This ``asymptotic
supersymmetry'' is a very puzzling phenomenon, whose full implications
are still not understood. Non supersymmetric tachyon free superstring
theories resemble models with spontaneously broken SUSY (of course
the scale of breaking is the Planck scale in general). Whether
this is more than a formal similarity remains to be seen. Note also
that the one loop cosmological constant in tachyon free string theories
is unnaturally small from the space-time point of view (although still
much too large). From our point of view, $\Omega$ can vanish without
space-time supersymmetry. This
requires only that $G_B$ and $G_F$ cancel precisely as $s\rightarrow0$.
We don't know whether or why this should be the case.

In theories with space-time fermions, one may also
define
$\tilde G(s)=G_B(s)+G_F(s)$, which counts the total density
of states of bosons {\it plus} fermions. This quantity
behaves when $s\rightarrow0$ as:
$\tilde G(s)\simeq s^a e^{b\over s}$,
and determines the thermal properties (e.g. the Hagedorn temperature)
of the theory.
It is not clear whether models with $b>0$
(such as those of section 3) exhibit any simplifying features
due to the fact that $G_B(s)-G_F(s)\simeq 0$ (as $s\rightarrow0$).

Our remaining task in this section is to state precisely and prove \pro.

\noindent{}{\bf Theorem:} Let $\zst(\tau)$ be a
modular invariant function, which is
finite throughout the fundamental domain $\CF$, except perhaps at
$\tau_2=\infty$, such that:

\noindent{}1) $\int_{\CF} {d^2\tau\over\tau_2^2}\zst={\rm finite}$. It is
implied in the definition of the above integral that for $\tau_2>>1$ we first
perform the integral over $-{1\over2}\leq\tau_1\leq{1\over2}$, and then
integrate over $\tau_2$.

\noindent{}2) As $\tau_2\rightarrow\infty$, $\zst\simeq \tau_2^x q^a\bar q^b$
($q=e^{2\pi i\tau}$) where $a,b>-1$.

Then:
\eqn\b{\int_{\CF}{d^2\tau\over\tau_2^2}Z_{\rm string}={\pi\over3}
\lim_{\tau_2\rightarrow0}\int_{-{1\over2}}^{1\over2} d\tau_1Z_{\rm string}
(\tau_1, \tau_2)}

\noindent{}{\bf Proof:}
Consider the function:
\eqn\fr{F(R)=\int_0^\infty{d\tau_2\over\tau_2^2}
\int_{-{1\over2}}^{1\over2}d\tau_1\zst(\tau)
\sum_{r\not=0}^\infty \exp\left({-\pi R^2r^2\over\tau_2}\right)}
This is a cutoff version of \sch, with a ``modular invariant''
cutoff. $R$ plays the role of $\Lambda^{-1}$ in the discussion above.
Our purpose is to relate the integral over the strip $F(R)$ \fr\ to
an integral over the fundamental domain $\CF$. Naively, this can be
done using
\ref\polc{J. Polchinski, Comm. Math. Phys. {\bf 104} (1986) 37;
B. Maclain and B. Roth, Comm. Math. Phys. {\bf 111} (1987) 539;
K. O' Brien and C. Tan, Phys. Rev. {\bf D36} (1987) 1184.}:
\eqn\naive{\int_{\CF}{d^2\tau\over\tau_2^2}\zst(\tau)
\exp\left(-\pi R^2{\vert n-m\tau\vert^2\over\tau_2}\right)
=\int_{\alpha\cdot \CF}{d^2\tau\over\tau_2^2}\zst(\tau)
\exp\left(-\pi R^2{r^2\over\tau_2}\right)}
where $\alpha$ is the  modular transformation taking
$e^{-\pi R^2{|n-m\tau|^2\over\tau_2}}\rightarrow e^{-\pi R^2{r^2\over\tau_2}}$.
We can write the sum over $r$ and integral over the strip \fr\ as a sum
over $r,\alpha$ and an integral over the fundamental domain $\CF$
(since the sum over $\alpha\cdot\CF$ generates the strip), and then
replace the sum over $r, \alpha$ by a sum over $n,m$ \naive\ (and integral
over $\CF$). This would suggest:
\eqn\true{\int_{\CF}{d^2\tau\over\tau_2^2}\zst(\tau)
\sum_{n,m\in Z}\exp\left(-\pi R^2{\vert n-m\tau\vert^2\over\tau_2}\right)
=F(R) + \int_{\CF}{d^2\tau\over\tau_2^2}\zst}
This is unfortunately too naive. The problem is that in many
interesting situations (most notably heterotic strings), the theory
contains ``unphysical tachyons'' (with $L_0\not=\bar L_0$),
which
lead to divergent terms as $q\rightarrow0$, but disappear after
the $\tau_1$ integration. In such situations the above integrals
(e.g. \naive) are not absolutely convergent. We have to integrate
the l.h.s. of \naive\ over $\tau_1$ first. In such cases, the
simple order of integration on $\CF$ translates to a complicated
prescription on $\alpha\cdot\CF$, different for each modular
transformation $\alpha$.
Since the integral depends on the order of integration, \true\ is in
general invalid. Clearly, in general the situation is out of control and the
argument of \polc\ can only be applied if we're dealing with
absolutely convergent integrals. Fortunately, this is the case
if $R$ in \true\ is large enough. By condition (2) of the theorem,
we are dealing with $\zst(\tau)$ such that there exists a $R_0$ such that
for all $R>R_0$ the l.h.s. of \naive\ is independent of the order of
integration, as long as $m\not=0$. But, since we use \naive\ only for
terms with $m\not=0$ (otherwise $\alpha=1$ in \naive), for $R>R_0$, \true\
is valid. Now we can continue \true\ analytically to $R<R_0$ and it
must still hold. The reason is that both the l.h.s. and the r.h.s.
of \true\ define analytic functions of $R$ in a strip ${\rm Im} R<
\epsilon,\; R>0$. For the l.h.s. this is clear, since the only
divergences come from $\tau_2\rightarrow\infty$, but by condition (2)
of the theorem no such divergence is possible for the function and
all its derivatives w.r.t. $R$. The r.h.s. is analytic since it is equal
to the l.h.s. for $R>R_0$, and the analytic continuation is unique.
We conclude that \true\ is correct al the way down to $R=0$. Poisson
resummation of the l.h.s. gives
\eqn\tr{{1\over R}\int_{\CF}{d^2\tau\over\tau_2^{3\over2}}\zst(\tau)
\sum_{n,m\in Z}q^{{1\over4}({n\over R}+mR)^2}\bar q^{{1\over4}({n\over R}-
mR)^2}=F(R)+\int_{\CF}{d^2\tau\over\tau_2^2}\zst(\tau)}
As $R\rightarrow0$ both sides behave as $1\over R^2$. Equating the
coefficients gives \b.

The conditions of the theorem are very natural. Condition 1 is obvious;
the main reason to require condition 2 is that it is equivalent
to unitarity when $c<c_{\rm critical}$. E.g. in \strings\
we have $\zst\simeq q^{\Delta-{c_M-1\over24}}\bar q^{\bar\Delta-
{c_M-1\over24}}$ (as $q\rightarrow0$), and as long as $c_M<25$,
if all $\Delta\geq0$ (as is the case in unitary CFT's),
condition 2 is satisfied. The reason one
expects the theorem to break down in general for non unitary matter,
is that then the Ramond sector may contain tachyons as well as the
NS sector. But in that case, bosonic and fermionic tachyons may cancel
in the expression for the free energy, giving rise to a finite
l.h.s. of \b, but still contribute to a large density of states,
such that the r.h.s. of \b\ is infinite. For $c>c_{\rm critical}$,
even unitary theories may violate the theorem. This region is not
understood, as mentioned above. Some subtleties may also occur
on the boundary between the two regions -- the critical (heterotic)
string (see \NADA\ for details).

\bigskip

\newsec{NON -- CRITICAL SUPERSTRINGS.}

In section 2 we saw that all non trivial theories of strings
without tachyon instabilities must contain space-time fermions,
in fact essentially the same number of fermions and of bosons. In the
critical (``ten dimensional'') case, this is achieved by a chiral GSO
projection \GSW. In this section we will see that a straightforward
generalization exists in the non-critical case as well, giving rise
to a large class of stable string theories with the number of degrees of
freedom
varying between that of two dimensional string theory and that of the critical
superstring. We will discuss the ``type II'' theory -- as usual there
exists a heterotic version.

Let us first recall how the GSO projection works in the case
of ten dimensional (perhaps compactified) string theory. One starts with
a fermionic string vacuum, which is normally left-right symmetric
(non chiral GSO projection), and therefore contains two sectors
(NS, NS) and (R, R) (for the left and right movers). Both sectors
contain bosonic excitations in space-time --
all states contribute with positive sign to the partition sum.
As discussed above, such theories are tachyonic due to the large number
of states ($c_{\rm string}>1$).
A chiral GSO projection can be implemented if the theory has a chiral $Z_2$
R -- symmetry. If there are no global anomalies, we can gauge
this $Z_2$ symmetry (or in other words orbifold by it \GSW); this
involves two elements; first eliminating all states which are not
invariant under the $Z_2$ symmetry from the spectrum. Hence only
a subset of the original bosonic (NS, NS) and (R, R) states survive
the projection. Then, as standard for orbifolds, there are also twisted
sectors, which in this context are the (NS, R) and (R, NS) sectors,
which contain space-time fermions. The theory thus obtained is not
necessarily space-time supersymmetric, however under favorable circumstances
it is tachyon free.

It is not known in general which $Z_2$ symmetries can be gauged and
furthermore give rise to tachyon free ``superstring'' models. There
is, however, a large class of theories, where this $Z_2$ is part of a
larger chiral algebra, which are in general tachyon free and possess
some additional nice properties. These are the space-time supersymmetric
string vacua. In general such vacua can be constructed iff the original
fermionic string vacuum has a global $N=2$ superconformal symmetry\foot{
To avoid confusion it is important to emphasize that this $N=2$
symmetry is an ``accidental'' global symmetry of the vacuum. It is not
gauged (the BRST charge is that of the $N=1$ fermionic string), and
most excitations are not invariant under this symmetry -- it is
a property of the vacuum and not of the full theory.
The existence of the $N=2$ symmetry is a necessary and sufficient
condition for space-time SUSY
\ref\FR{D. Friedan, A. Kent, S. Shenker and E. Witten, unpublished;
T. Banks, L. Dixon, D. Friedan and E. Martinec, Nucl. Phys.
{\bf B299} (1988) 613.}.}.
The projection can then be performed by constructing the (chiral)
space-time SUSY charge $S$
\ref\FMS{D. Friedan, E. Martinec and S. Shenker, Nucl. Phys.
{\bf B271} (1986) 93.}, and projecting out all operators which are not
local w.r.t. $S$; then one should again add ``twisted sectors''
obtained by acting with $S$ on the remaining (NS, NS) and (R, R)
operators; this gives rise to the (R, NS) and (NS, R) sectors (space-time
fermions). This procedure is expected to be in general free of global
anomalies (modular invariant), although no general proof exists.
Also, such vacua are automatically free of tachyons, assuming
there is only one time direction and the remaining ``matter'' is unitary.
This follows from the space-time SUSY algebra. We will not
elaborate on this procedure here, but rather explain directly how
it generalizes to the non-critical case. For details on the ``critical''
construction we refer the reader to the original literature \FMS, \FR,
\ref\gepner{D. Gepner, Nucl. Phys. {\bf B296} (1988) 757;
PUPT-1121 (1989).}.

The procedure we are going to describe turns
(non-critical) fermionic string vacua
with an accidental global $N=2$ symmetry to consistent superstrings.
We will describe it in the particular case of the $D=d+1$ dimensional
fermionic string. The general structure is an obvious generalization,
and can be found in \KUS.

The $d+1$ dimensional fermionic string is described in superconformal
gauge by $d$ matter superfields $X_i$, $i=1,...,d$, and the super
Liouville field $\Phi$. In components we have:
\eqn\supF{\eqalign{
X_i=&x_i+\theta\psi_i+\bar\theta\bar\psi_i+\theta\bar\theta F_i;
\; i=1,...,d\cr
\Phi=&\phi_l+\theta\psi_l+\bar\theta\bar\psi_l+\theta\bar\theta F_l\cr}}
The fields $X_i$ are free, while $\Phi$ is described by the super Liouville
Largangian \POL:
\eqn\SLaction{S_{SL}={1\over2\pi}\int d^2z\int d^2\theta\left[
D\Phi\bar D\Phi+
2\mu\exp(\alpha_+\Phi)\right]}
where $D=\partial_\theta+\theta\partial_z$, and we have dropped
curvature couplings \SL.
The central charge is $\hat c_L(={2\over3} c_L)=1+2Q^2$,
$Q=\sqrt{9-d\over2}$.
As mentioned above, this theory has a tachyonic ground state (for
$d>1$). The corresponding vertex operator is
\eqn\tachyon{T_k=\int d^2\theta \exp(ik\cdot X+\beta\Phi)}
where ${1\over2}k^2-{1\over2}\beta(\beta+Q)={1\over2}$. As in the bosonic
case, $E=\beta+{Q\over2}$ (see section 1 \ps), $E^2-k^2=m^2={1-d\over8}$,
so that $T_k$ is a tachyon for $d>1$.

For $d=9$ (critical ten dimensional
string theory), the $Z_2$ R -- symmetry that one orbifolds by to obtain
the superstring is $\psi\rightarrow-\psi$, $\bar\psi\rightarrow\bar
\psi$ (or equivalently $\theta\rightarrow-\theta, \;\bar\theta\rightarrow
\bar\theta$). The tachyon \tachyon\ transforms to minus itself under
this symmetry, and is projected out of the spectrum.
It is tempting to try and divide by the same symmetry for all $d$.
Unfortunately, this is too naive; for $d<9$ the orbifoldized theory
doesn't make sense; there are a number of ways of seeing that --
modular invariance breaks down, the spin field \FMS\ doesn't have the
necessary properties, etc. We have to use the more sophisticated approach
related to global $N=2$ symmetry.
The basic observation is that for odd $d=2n+1$, such that the total dimension
of space-time is even\foot{There is an analogous requirement
in compactified critical string theory.}, the system \supF\ possesses
an $N=2$ symmetry. This symmetry pairs $d-1=2n$ of the $X_i$ into $N=2$
superfields, and the remaining $X_{2n+1}\equiv X$ is paired with $\Phi$.
This involves breaking the $O(d)$ symmetry; one can think of
$\Phi, X$ as light cone coordinates, while the rest of the $X_i$
($i=1,...,2n$) are transverse directions.
The vacuum is not translationally invariant in the $\Phi$ direction.
The action of the $N=2$ superconformal symmetry on the
transverse $X_i$ is completely standard and we will not review it.
The free $\Phi, X$ system forms a chiral $N=2$ multiplet:
it is convenient to define
$\phi=\phi_l+ix,\;\psi=\psi_L+i\psi_x$, etc. The $N=2$ generators
are:
\eqn\one{\eqalign{T(z)=&-{1\over2}\partial\phi\partial\phi^\dagger+{1\over4}
(\psi\partial\psi^\dagger+\psi^\dagger
\partial\psi)-{1\over4\gamma}\partial^2\phi
-{1\over4\gamma}\partial^2\phi^\dagger\cr
J(z)=&{1\over2}\psi\psi^\dagger
-{1\over2\gamma}\partial\phi+{1\over2\gamma}\partial\phi^\dagger
\cr
T_F^+=&{1\over2}\psi^\dagger\partial\phi+{1\over2\gamma}\partial\psi^\dagger\cr
T_F^-=&{1\over2}\psi\partial\phi^\dagger+{1\over2\gamma}\partial\psi\cr}}
In \one, $\dagger$ does not interchange left and right movers -- it corresponds
to complex conjugation in field space:
$\phi^\dagger=\phi_l-ix$, etc.
As in Liouville theory, the free field expressions \one\ should be
understood literally away from an interacting region (``wall''),
which we haven't
specified yet. The string coupling is again (as in the bosonic
theories) $g_{\rm st}\propto e^{-{Q\over2}\phi}$, and we need a potential
to suppress the region $\phi\rightarrow-\infty$; we haven't yet determined
the appropriate potential. The form \one\ is still very useful, for example
to discuss bulk effects, which are anyway the best understood part
of the Liouville correlation functions.

To choose an appropriate wall, it is convenient to define the
chiral $N=2$ superfield $\Gamma=\phi+\theta\psi+\bar\theta\bar\psi+
\theta\bar\theta F+\cdots$ (such that
$D^\dagger\Gamma=0$). Here $\theta$ is a complex Grassmanian variable
(different from the $\theta$ in \supF\ -- \tachyon). In terms of $\Gamma$, the
free Largrangian for $\Phi, X$ is:
$${\cal L}={1\over8\pi}\int d^4\theta\Gamma\Gamma^\dagger$$
We would like now an $N=2$ invariant operator to provide a potential for
$\phi$ and to set the scale.
There is a very natural choice for the potential, which preserves
the $N=2$ symmetry:
\eqn\three{{\cal L}={1\over8\pi}\int d^4\theta
\Gamma^\dagger\Gamma+i{\mu\over 8\pi \gamma} \int d^2 \theta e^{-\gamma\Gamma}+
i{\mu\over 8 \pi\gamma} \int d^2 \theta^\dagger e^{-\gamma\Gamma^\dagger}
}
The potential is an F-term, while from the $N=1$ point of view it is
a tachyon condensate at some particular non zero momentum. This
value of the momentum is special because \three\ is actually manifestly
$N=2$ supersymmetric; in fact, it is precisely the $N=2$ Liouville
Lagrangian. However its role here is to supply an accidental
global symmetry -- it is {\it not}
a remnant of $N=2$ supergravity in superconformal
gauge.
The parameter $\gamma$ is not renormalized in $N=2$ Liouville
(unlike in $N=0,1$
Liouville): $\gamma={1\over Q}$. The $N=2$ Liouville system \three\
together with the $2n$ $X_i$
forms a vacuum of ($N=1$) fermionic string theory with a global
$N=2$ symmetry. However, this is not the superstring yet -- the theory
still contains tachyons \tachyon, does not contain space-time fermions, etc.
To perform the chiral GSO projection, we define the target space SUSY
operator $S$:
\eqn\Sz{S(z)=e^{-{1\over2}\sigma+{i\over2}(H_1+...+H_n+H_l+Qx)}}
Here $\sigma$ is the bosonized superconformal ghost current \FMS:
$\beta\gamma=\partial\sigma$; the $H_i$ are bosonized fermions:
$\psi_{2i-1}\psi_{2i}=\partial H_i$ ($i=1,...,n$), and similarly
$\psi_l\psi_x=\partial H_l$. The GSO projection amounts to keeping
in the spectrum only operators which have a local OPE with $S(z)$.
Note that apart from the last term ($e^{{i\over2}Qx}$) in $S$, this
just means projecting by $(-)^{F_R}$, where $F_R$ is the right moving
fermion number (the naive projection $\psi\rightarrow-\psi$).
For $Q=0$ (the critical case) we recover the original GSO projection.
In general, the projection ties the momentum in the $x$ direction
with the fermion number. In particular, $p_x$ takes discrete values.
The operator $S(z)$ is chiral ($\bar\partial S=0$); this seems strange
in view of the fact that $x$ can be non-compact (so that only operators
with $p_{\rm left}=p_{\rm right}$ are physical). The point is that the
spectrum of $x$ momenta is determined by locality w.r.t. $S$; after
the projection, the radius of $x$ is not a meaningful concept.

After choosing the ``longitudinal'' direction, the theory still
has $SO(2n)$ rotation invariance in the $X_i$ directions. This means
that the supercharge $S$ \Sz\ is not unique. Additional
charges may be obtained by acting on it with the $SO(2n)$ generators
$e^{\pm iH_a\pm iH_b},\; (a\not=b=1,2,...,n)$. This gives a set of charges
$S_\alpha$ in one of the spinor representations of $SO(2n)$. To get
the other representation (for $n>0$), one uses the fact that the operator:
\eqn\tdS{{\tilde S}(z)=e^{-{1\over2}\sigma+{i\over2}
(-H_1-...-H_{n-1}+H_n-H_l-Qx)}}
is local w.r.t. $S$, and therefore is physical. Applying $SO(2n)$
rotations to $\tilde S$ generates the second spinor representation. For
odd $n$ the two are isomorphic, while for even $n$, $S_\alpha$ and
$S_{\dot \beta}$ transform as different representations. The zero modes:
\eqn\zerom{Q_\alpha=\oint{dz\over2\pi i} S_\alpha(z);\;
\tilde Q_{\dot\beta}=\oint{dz\over2\pi i}\tilde S_\beta(z)}
satisfy the super-algebra
\eqn\supals{\eqalign{
\{ Q_\alpha, \tilde Q_{\dot \beta}\} &= \gamma^i_{\alpha \dot \beta} P_i
\qquad\qquad {\rm for} \;n \;{\rm even} \cr
\{ Q_\alpha, \tilde Q_{\beta}\} &= \gamma^i_{\alpha \beta} P_i
\qquad\qquad {\rm for} \;n \;{\rm odd} \cr}
}
The algebra \supals\ is a ``space SUSY'' algebra in the transverse
directions:
lightcone ($\phi_l, x$) translations do not appear on the r.h.s., essentially
because the vacuum is not translationally invariant in $\phi$.

We now turn to discuss some general features of the spectrum of these
theories. Although most states are paired by the SUSY algebra \supals,
apriori some tachyonic states could remain, since lack of tachyons
is no longer implied by the algebra. However, for unitary matter
theories (such as \supF), one can check directly in general that the
theory is tachyon free after the GSO projection. From the point of view
of section 2 this is clear: the effective density
of states of the GSO projected theory
\gs\ receives contributions only from states which are killed by
the supercharges:
\eqn\cancel{Q_\alpha|{\rm phys}\rangle=\tilde Q_{\dot\beta}|{\rm phys}\rangle
=0}
The rest of the states are paired level by level and cancel in the
free energy \gs. Now \cancel\ implies that $|{\rm phys}\rangle$
is independent of the transverse excitations (more generally of the
transverse excitations and the internal degrees of freedom); the number
of states that can satisfy \cancel\ is therefore of the order
of the number of states in the $\Phi, X$ (or $\Gamma$) system, which
is a 2D string system. Hence $\lim_{s\rightarrow0} G(s)$ is finite,
and as shown in section 2, tachyons can not exist by modular invariance.
Unitarity enters in two places:

\noindent{}1) Condition (2) of the theorem proved in section 2 breaks
down in general in non unitary theories.

\noindent{}2) If the conditions are satisfied,
the argument above proves only that the one loop
partition sum is finite. There could be cancellations between NS
and R tachyons, leading to a finite partition sum in the presence
of tachyons. In unitary theories, Ramond tachyons can not occur
(even before the chiral projection). Hence, only (NS, NS) tachyons
may exist, and these are ruled out by the above argument.

Absense of tachyons can also be established directly \KUS, but we feel
that the above argument is more intuitive. In the $D$ dimensional
superstring the explicit check of absense of tachyons is particularly
simple: the only tachyon before the projection is $T_k$ \tachyon.
The GSO projection (locality of $T_k$ w.r.t. $S$ \Sz) implies:
$k_xQ\in 2Z+1$. The smallest value of $k_x$ (corresponding to the
lightest state) is $|k_x|={1\over Q}$, for which $(\beta+{Q\over2})^2=
{Q^2\over4}+{1\over Q^2}-1\geq0$ for all $Q$. Hence the momentum in the
$x$ direction is quantized precisely such that the tachyon is always
massive. In particular, the zero momentum tachyon (the
original cosmological
term in \SLaction) is projected out\foot{Note also that the remaining
modes of the tachyon go to infinite mass as $Q\rightarrow0$.
Therefore, they are invisible in the critical theory.}.

Another interesting property of these theories is that target SUSY
can be broken continuously, by switching on the modulus
$\lambda\int\partial x\bar\partial x$. This perturbation
is not projected out by GSO; it is clear from the form
\Sz\ of the generators that the symmetry is broken for any $\lambda\not=0$.
The situation is quite different from the ten dimensional
case, where continuous breaking of SUSY can not occur
\FR. The difference is due to the time dependence of the
vacuum solution, but the detailed mechanism is still unclear; the
gravitino is massive here even when SUSY is unbroken.

Returning to the $N=2$ Liouville action, one can ask what is the role
of the $N=2$ cosmological term: why did we choose an $N=2$ invariant
perturbation to set the scale, out of the infinite number
of GSO invariant operators, which are not $N=2$ invariant (in general).
After all, the analysis of the spectrum and symmetries are in any
case performed in the region where the interaction is negligible.
The answer is that it is indeed admissible to set the scale with other
on shell physical vertex operators. Nevertheless, we do expect
the $N=2$ invariant perturbation \three\ to exhibit especially
nice properties. First, since \three\ is $N=2$ invariant, we have manifest
space SUSY for all $\mu$. For other perturbations in \three, SUSY
would in general be broken. In addition, F-terms in $(2,2)$ theories
are known to yield moduli of the
appropriate CFT (see e.g.
\ref\dixon{L. Dixon, Princeton preprint PUPT-1074 (1987).}).
For generic perturbations, the dynamics of the corresponding ``Liouville''
model \three\ is probably much more complicated.

The $N=2$ cosmological constant has $\beta=-\gamma=-{1\over Q}$,
or $E=\beta+{Q\over2}={Q\over2}-{1\over Q}$.
This is positive only when $Q^2>2$. As $Q^2\rightarrow2$ $(d\rightarrow5)$,
the lowest lying NS state goes to zero mass, and for $d>5$, it energy
becomes negative (while the mass
increases). As menioned above, the behavior of $E>0$ and
$E<0$ states is qualitatively different. One would need a more detailed
understanding of the theory to see what changes as $E\rightarrow0$, but it
is clear that the point $Q^2=2$ corresponds to some kind of transition
in the behavior of the theory. For $Q^2<2$ there are other possibilities
to set the scale in \three\ with operators of positive energy, however
they are all less symmetric.
The special role of this new ``critical dimension'' is not entirely clear,
and should be elucidated further.

The GSO procedure may superficially resemble fine tuning
the coefficients of all the tachyonic modes in the action to zero.
The difference between this and fine tuning is that in the GSO
procedure the states are projected out by gauge invariance
(the $Z_2$ R symmetry). Therefore,
we are assured that they will not reappear in various intermediate channels
in higher loops. On the other hand, as we saw in section 2, fine tuning
does not solve the problem of higher loops (topologies).

An interesting property of the non-critical superstrings we have constructed
is that they have a vanishing partition sum (generically) to all orders
in the genus expansion. Consider first the partition sum on the sphere:
usually, the spherical partition sum is not zero in non-critical string theory
\ref\zamw{A.B. Zamolodchikov, \pl {\bf 117B} (1982) 87.}.
The reason is the following. In critical string theory, the partition
sum vanishes due to the six $c, \bar c$ (reparametrization ghost)
zero modes on the sphere. However, in the non-critical case, the
Liouville path integral contains a comparable divergence, due to
zero modes of the classical Liouville solution. The zero related to the
$c,\bar c$ zero modes is really $1\over {\rm vol} SL(2, C)$, and the infinity
in the Liouville sector is proportional to the same volume (up to a finite
factor), thus the two cancel. More precisely, choosing the conformal gauge
does not fix the gauge completely. By fixing the $SL(2, C)$ invariance
and therefore not integrating over the various zero modes, one finds
a finite answer. In fermionic string theory a similar phenomenon
occurs, with $SL(2, C)$ replaced by $OSp(2,1)$. Again, the infinite
factors cancel, and one is left with a finite partition sum.
In the limit $Q\rightarrow0$ (when the non-critical string approaches
a critical one), we still have a finite partition sum, but exactly
at $Q=0$, $\phi$ translation invariance appears, and we have to
divide this finite answer by the volume of $\phi$. The free
energy per unit volume
${{\rm const}\over V}\rightarrow0$ as $V\rightarrow\infty$ ($\phi$ is
non compact). Thus, in space-time the difference between the critical and the
non-critical cases is that in the latter, due to lack of $\phi$ translation
invariance, we compute the total free energy, which is finite, while
in the critical case,
where translational invariance is restored,
we are interested in the free energy per unit volume,
which is zero.

When the fermionic string vacuum possesses a global $N=2$ symmetry
(even before the chiral projection, which is of course irrelevant
for the partition sum on the sphere), there are {\it two additional} Liouville
fermion zero modes, obtained by applying $N=2$ transformations to the
usual fermionic zero modes, which exist for $N=1$ Liouville. The ghosts,
which are still those of the $N=1$ string do not have balancing zero
modes, therefore the path integral vanishes. This means that the classical
vacuum energy vanishes in \three. The torus, and higher genus
partition sums also vanish for $n\geq1$, by the usual contour
deformation arguments \FMS,
\ref\emil{E. Martinec, \pl {\bf 171B} (1986) 189.}.
In that case, bosons and fermions are paired, except perhaps
for a set of measure zero
of states at zero momentum $p_i=0$ ($i=1,...,2n$). Sometimes, such arguments
can be subtle due to contact terms from boundaries of moduli
space, however here we expect such subtleties to be absent in general, since
there are no massless excitations.

To illustrate the above abstract discussion, we finish this section
with a brief analysis of the simplest theories constructed here: the
two and four dimensional superstrings ($n=0,1$).

\medskip

{\bf Example 1: Two dimensional superstring\foot{\rm Based on
\ref\ND{D. Kutasov, G. Moore and N. Seiberg, unpublished.}.}.}

This theory contains two superfields: the super Liouville field
$\Phi$, and a space coordinate $X$; the two combine into an $N=2$ Liouville
system \one\ with $\gamma(={1\over Q})={1\over2}$.
To calculate the torus partition sum, we have to sum \Zt\ over all
states satisfying:

\noindent{}(a) Locality w.r.t. \Sz; here, $S(z)=e^{-{1\over2}\sigma+
{i\over2}H+iX}$.

\noindent{}(b) $\Delta-\bar\Delta\in Z$.

\noindent{}(c) Mutual locality.

\noindent{}There are four sectors in the theory. We will next go over
them and solve the conditions (a) -- (c).

\noindent{}Consider first the (NS, NS) sector:
the vertex operators have the form
$$e^{-\sigma+inH_l+ipX}
e^{-\bar\sigma+i\bar n\bar H_l+i\bar p\bar X}$$
Condition (a) leads to
\eqn\conda{\eqalign{
p=&{n-1\over2}+m\cr
\bar p=&{\bar n-1\over2}+\bar m\cr}}
$(m, \bar m\in Z)$. From condition (b) we have
\eqn\condb{\eqalign{n-\bar n&\in 2Z\cr
p^2-\bar p^2&\in 2Z\cr}}
while from condition (c):
\eqn\condc{pp^\prime-\bar p\bar p^\prime\in Z}
for all $p, p^\prime$ in the (NS, NS) spectrum. The solution of
the constraints \conda\ --
\condc\ is:

\noindent{}(1) $n, \bar n\in 2Z+1;\; p=m,\; \bar p=\bar m,\;m-\bar m\in
2Z$.

\noindent{}(2) $n, \bar n\in 2Z;\; p=m+{1\over2},\;
\bar p=\bar m+{1\over2},\;m-\bar m\in
2Z$.

Summing \one\ over these states gives (after multiplying by the
oscillator contribution for $X$, $H_l$, and the ghost contribution):
\eqn\NSNS{Z_{\rm NS, NS}={\sqrt{\tau_2}\over2}|\theta_2|^2}
A similar analysis applied to the other 3 sectors gives
\eqn\NSR{Z_{\rm NS, R}=Z_{R, NS}=-{\sqrt\tau_2\over4}|\theta_2|^2}
\eqn\RR{Z_{\rm R, R}=\sqrt{\tau_2}\left[|\theta_2|^2+|\theta_3|^2+
|\theta_4|^2\right]}
To see better where the different contributions are coming from,
we should list the physical operators which survive the projection.
Chirally, in the NS sector we have the tachyon operators:
\eqn\projtach{T_k=e^{-\sigma+ikX+\beta\phi_l};\; k\in Z+{1\over2}}
The restriction on the spectrum of $k$'s is due to the requirement
of locality with $S(Z)$ \Sz.
In the Ramond sector we have two sets of operators of the form:
\eqn\projRam{V_k=e^{-{1\over2}\sigma+{i\over2}\epsilon H+ikX+\beta\phi_l}}
with
\eqn\Rspec{\eqalign{
(1)\;\;\epsilon=1&\;\;\;k=n=0,1,2,3,\cdots;\;\;\; \beta=n-1\cr
(2)\;\;\epsilon=-1&\;\;\;k=-l-{1\over2}\;\;\;l=0,1,2,3,\cdots;\; \;\;\beta=
l-{1\over2}\cr}}
In agreement with the partition sums \NSNS\ -- \RR. Due to the low
space-time dimension, the superalgebra \supals\ is quite degenerate: there
is only one supercharge, $Q=\oint{dz\over2\pi i}S(z)$, \Sz, which satisfies
$Q^2=0$. Hence, $Q$ is a BRST like charge. An interesting theory
is obtained by restricting ourselves to ${\rm ker}\; Q$:
physical operators are those which satisfy:
\eqn\Qphys{Q|{\rm phys}\rangle=0}
This gives a topological string theory (reminiscent of the
topological model of the bosonic string with $c=-2$ matter
\ref\TOP{
E. Witten, Nucl. Phys. {\bf B340} (1990) 281;
J. Distler, Nucl. Phys. {\bf B342} (1990) 523.}). The condition \Qphys\
projects out the tachyon modes $T_k$ \projtach\ with $k<0$, and the
second set of Ramond states in \Rspec\ (those with $\epsilon=-1,\;k<0$).
Since the physical states must have the same Liouville momentum for left
and right movers ($\phi_l$ is non compact), all space-time fermions
are projected out by \Qphys. The topological two dimensional superstring
contains (NS, NS) tachyons
\eqn\Tk{T_n=e^{-\sigma+i(n+{1\over2})x+(n-{1\over2})\phi_l};\;n=0,1,2,...}
and (R, R) states:
\eqn\RRs{V_n=e^{-{1\over2}\sigma+{i\over2}H_l+
inx+(n-1)\phi_l};\;n=0,1,2,...}
(here we mean left-right symmetric combinations $\bar V_{\rm left}
V_{\rm right}$).
Correlation functions of the operators \Tk, \RRs\ can be obtained
using the methods of
\ref\DIIFK{P. Di Francesco and D. Kutasov, Princeton preprint PUPT-1276
(1991).}.
We leave further investigation of this topological theory to future
work.

If one does not impose \Qphys\ on the spectrum, one can ask how does
the theory change when we add the (physical) operator $\partial x
\bar\partial x$ to the action. The dependence
of the partition sum on the ``radius'' of $x$
can be obtained
by viewing the theory
as a chiral orbifold. The idea is the following: if $x$ has radius
$R$ ($x\simeq x+2\pi R$), we can study the orbifold theory
obtained by gauging the $Z_2$ symmetry:
\eqn\orb{\eqalign{x\rightarrow&x+\pi R\cr
H_l\rightarrow&H_l+\pi\cr
\bar H_l\rightarrow& \bar H_l\cr
\sigma\rightarrow&\sigma-i\pi\cr
\bar\sigma\rightarrow&\bar\sigma+2\pi i\cr
}}
This formulation of the theory as a chiral orbifold emphasizes the role
of the $Z_2$ symmetry which is being gauged \orb. The role of the space SUSY
operator $S(z)$ is obscure from this point of view. In particular, the symmetry
generator $Q$ \zerom\ is in the spectrum only for particular $R$ ($R=2$).
This symmetric point is not singled out in \orb.

By standard methods one obtains the partition sums:
\eqn\NSNSO{Z_{NS, NS}(R)=\sqrt{\tau_2}\sum_{m,m^\prime\in Z}
q^{{1\over2}({2m+1\over R}+{m^\prime R\over2})^2}\bar q^{{1\over2}(
{2m+1\over R}-{m^\prime R\over2})^2}}
\eqn\RRO{Z_{R, R}(R)=\sqrt{\tau_2}\sum_{m,m^\prime\in Z}
q^{{1\over2}({m\over R}+{m^\prime R\over2})^2}\bar q^{{1\over2}(
{m\over R}-{m^\prime R\over2})^2}}
\eqn\RNSO{Z_{NS, R}(R)+Z_{R, NS}(R)=-\sqrt{\tau_2}\sum_{m,m^\prime\in Z}
q^{{1\over2}({2m\over R}+{R\over2}(m^\prime+{1\over2})
)^2}\bar q^{{1\over2}(
{2m\over R}-{R\over2}(m^\prime+{1\over2}))^2}}
For $R=2$ we have the results \NSNS, \NSR, \RR. For generic $R$, the total
partition sum can be written as a linear combination of $c=1$ partition sums:
$Z_{\rm total}(R)=2Z_{c=1}(R)-Z_{c=1}({4\over R})$.
The integral $\Omega=\int_{\CF}{d^2\tau\over\tau_2^2}Z_{\rm total}$,
which as we have learnt in section two gives the number of degrees of freedom,
can be performed using \b\ (or the results of
\ref\oneloop{M. Bershadsky and I. Klebanov, Phys. Rev. Lett. {\bf 65}
(1990) 3088; N. Sakai and Y. Tanii, Tokyo preprint TIT/HEP-160 (1990).});
one finds that $\Omega\propto R$. In particular, at the ``supersymmetric''
, or topological, point $R=2$, the cancellation of bosons and fermions
is not complete. This is due to the low space-time dimension -- the SUSY
algebra $Q^2=0$ does not imply pairing of bosons and fermions. As explained
above, we need at least two non compact transverse directions for that.
Therefore, we will next consider
the case of two transverse dimensions.

\medskip

{\bf Example 2: Four dimensional superstring.}

The ``matter'' consists of two superfields
$X_1, X_2$ (in addition to $\Phi, X$), $Q=\sqrt3$, and
$S=e^{-{1\over2}\sigma+{i\over2}(H_l+H_1+\sqrt{3}x)}$. One can repeat
the discussion of Example 1 here;
we will leave the details to the reader, and only give the final
result for the one loop partition sums\foot{All sums over $\epsilon$ run
over $\epsilon=0,2,4$.}:
\eqn\NSNSfour{\eqalign{
\sqrt{\tau_2}|\eta(\tau)|^6Z_{NS, NS}=&|\theta_3+\theta_4|^2
\sum_\epsilon\Theta\left[{{\epsilon+3\over6}\atop 0}\right](12\tau)
\bar\Theta\left[{{\epsilon+3\over6}\atop0}\right](12\bar\tau)+\cr
&|\theta_3-\theta_4|^2\sum_\epsilon\Theta\left[{{\epsilon\over6}\atop0}
\right](12\tau)\bar\Theta\left[{{\epsilon\over6}\atop0}\right](12\bar\tau)+\cr
&(\theta_3+\theta_4)(\bar\theta_3-\bar\theta_4)\sum_\epsilon
\Theta\left[{{\epsilon+3\over6}\atop0}\right](12\tau)\bar\Theta
\left[{{\epsilon\over6}\atop 0}\right](12\bar\tau)+\cr
&(\theta_3-\theta_4)(\bar\theta_3+\bar\theta_4)\sum_\epsilon\Theta\left[
{{\epsilon\over6}\atop0}\right](12\tau)\bar\Theta\left[{{\epsilon+3\over6}
\atop0}\right](12\bar\tau)\cr}}
\eqn\RRfour{\eqalign{
\sqrt{\tau_2}|\eta(\tau)|^6Z_{R,R}=&|\theta_2|^2
\sum_\epsilon
\Theta\left[{{\epsilon-{1\over2}\over6}\atop0}\right](12\tau)\bar\Theta
\left[{{\epsilon-{1\over2}\over6}\atop0}\right](12\bar\tau)+\cr
&|\theta_2|^2\sum_\epsilon
\Theta\left[{{\epsilon+3-{1\over2}\over6}\atop0}\right](12\tau)\bar\Theta
\left[{{\epsilon+3-{1\over2}\over6}\atop0}\right](12\bar\tau)+\cr
&|\theta_2|^2\sum_\epsilon
\Theta\left[{{\epsilon-{1\over2}\over6}\atop0}\right](12\tau)\bar\Theta
\left[{{\epsilon+3-{1\over2}\over6}\atop0}\right](12\bar\tau)+\cr
&|\theta_2|^2\sum_\epsilon
\Theta\left[{{\epsilon+3-{1\over2}\over6}\atop0}\right](12\tau)\bar\Theta
\left[{{\epsilon-{1\over2}\over6}\atop0}\right](12\bar\tau)+\cr
}}
\eqn\NSRfour{\eqalign{
-\sqrt{\tau_2}|\eta(\tau)|^6Z_{NS, R}=&(\theta_3+\theta_4)\bar\theta_2
\sum_\epsilon\Theta\left[{{\epsilon+3\over6}\atop0}\right](12\tau)
\bar\Theta\left[{{\epsilon+{3\over2}\over6}\atop0}\right](12\bar\tau)+\cr
&(\theta_3-\theta_4)\bar\theta_2
\sum_\epsilon\Theta\left[{{\epsilon\over6}\atop0}\right](12\tau)
\bar\Theta\left[{{\epsilon-{3\over2}\over6}\atop0}\right](12\bar\tau)+\cr
&(\theta_3+\theta_4)\bar\theta_2
\sum_\epsilon\Theta\left[{{\epsilon+3\over6}\atop0}\right](12\tau)
\bar\Theta\left[{{\epsilon-{3\over2}\over6}\atop0}\right](12\bar\tau)+\cr
&(\theta_3-\theta_4)\bar\theta_2
\sum_\epsilon\Theta\left[{{\epsilon\over6}\atop0}\right](12\tau)
\bar\Theta\left[{{\epsilon+{3\over2}\over6}\atop0}\right](12\bar\tau)+\cr
}}
Now, by general arguments above, we expect $Z_{\rm total}=Z_{NS,NS}+
Z_{R,R}+Z_{NS,R}+Z_{R, NS}$ to vanish!
Quite miraculously from the
point of view of the explicit formulae \NSNSfour\ -- \NSRfour, one
can write $Z_{\rm total}$ in a much simpler way as:
\eqn\simpleZ{\sqrt{\tau_2}|\eta(\tau)|^6Z_{\rm total}=\sum_\epsilon
|F_\epsilon(\tau)|^2}
where
\eqn\Fe{\eqalign{
F_\epsilon(\tau)=&(\theta_3+\theta_4)(\tau)\Theta\left[{{
\epsilon+3\over6}\atop0}\right](12\tau)+(\theta_3-\theta_4)(\tau)\Theta
\left[{{\epsilon\over6}\atop0}\right](12\tau)\cr
-&\theta_2(\tau)\left\{
\Theta\left[{{\epsilon+{3\over2}\over6}\atop0}\right](12\tau)+
\Theta\left[{{\epsilon-{3\over2}\over6}\atop0}\right](12\tau)\right\}\cr}}
The most natural scenario would be to have $F_\epsilon(\tau)=0$
for all $\epsilon(=0,2,4)$. This seems indeed to be the case. We have checked
the vanishing of $F_\epsilon$ \Fe\ for low
orders in $q$, and believe that the result is true in general,
although we haven't proved it analytically. If this is the case, we have, as
expected, $Z_{\rm total}=0$.

\bigskip

\newsec{CLASSICAL DYNAMICS IN 2D STRING THEORY.}

In the previous sections we have seen that stable
(tachyon free) string vacua are in
general those with $c_{\rm string}=1$. Bosonic strings with $c_{\rm string}
=1$ (two dimensional strings), are known from the matrix models
\ORIG, \CONE,
\ref\MOORE{G. Moore,
Rutgers preprint RU-91-12 (1991).},
\ref\INTER{D. Gross and I. Klebanov, Princeton preprint PUPT-1241 (1991);
G. Mandal, A. Sengupta, and S. Wadia, IAS preprint, IASSNS-HEP/91/8 (1991);
K. Demeterfi, A. Jevicki, and J.P. Rodrigues, Brown preprints BROWN-HET-795,
803
(1991); J. Polchinski, Texas preprint UTTG-06-91 (1991).}\
to be solvable. In this section we will review
the understanding of their properties in the continuum path integral
formalism, in the hope that some of them may carry over
and help to understand more complicated theories with $c_{\rm string}=1$,
like those of section 3.

In the introduction, we have seen that the set of ``resonant'' amplitudes
defines for all $d$ a ``bulk'' S -- matrix, which is given by Shapiro --
Virasoro type integral representations. To illustrate this, we discuss here
tachyon dynamics in $D=d+1$ dimensional ``non-critical'' string theory.
Of course, for generic $D$, there is no reason to concentrate on the tachyon,
both because it is merely the lowest lying state of the string spectrum,
and because it is tachyonic, thus absent in more physical theories (see
section 3). Our justification is that for $D=2$ it is the only field theoretic
degree of freedom, and is massless.

The on shell vertex operator for the tachyon is:
\eqn\tach{T_k=\exp(ik\cdot X+\beta(k)\phi)}
where $k, X$ are $d$ -- vectors, and by \mass:
\eqn\mass{{1\over2}k^2-{1\over2}\beta(\beta+Q)=1}
which implies as before \dmass:
$m_{\rm tachyon}^2={2-D\over12}$. The bulk amplitudes
described in the introduction take here the form:
\eqn\tachcor{A_{s=0}(k_1,..k_N)=\langle T_{k_1}..T_{k_N}\rangle/\log\mu}
where $\sum_ik_i=0$, and by \Spar\ $\sum_i\beta_i=-Q$.
As mentioned above, such amplitudes have an integral representation
for all $D$:
\eqn\Npoint{A_{s=0}(k_1,..,k_N)=\prod_{i=4}^N
\int d^2z_i\vert z_i\vert^{2(k_1\cdot
k_i-\beta_1\beta_i)}\vert 1-z_i\vert^{2(k_3\cdot k_i-\beta_3\beta_i)}
\prod_{4=i<j}^N\vert z_i-z_j\vert^{2(k_i\cdot k_j-\beta_i\beta_j)}}
For $D=26$ this is the familiar Shapiro -- Virasoro representation \GSW,
however it retains all the nice properties for all $D$. The singularities,
which are the most important features of \Npoint, are poles in various
channels, corresponding to on shell intermediate particles. To examine
these poles, one focuses on contributions to \Npoint\ of regions
in moduli space where vertices approach each other. E.g. from the
region $z_4, z_5,..., z_{n+2}\rightarrow0$, we have an infinite
series of poles in $(E,p)$, $E={Q\over2}+\sum_i\beta_i$, $p=\sum_i k_i$
($i=1,4,5,6,...,n+2$), with residues related to lower correlation
functions with insertions of various intermediate states. For
example, near the first (tachyon) pole we have:
\eqn\resN{A_{s=0}(k_1,..,k_N)\simeq
{\langle T_{k_1}T_{k_4}...T_{k_{n+2}}T_{\tilde k}\rangle
\langle T_{\Sigma_i k_i}T_{k_2}T_{k_3}T_{k_{n+3}}.. T_{k_N}\rangle
\over ({Q\over2}+\sum_i\beta_i)^2
-(\sum_ik_i)^2-{2-D\over12}}}
where $\tilde k=-\sum_ik_i$, and similarly for the higher poles.
Thus, for generic correlators \Npoint, the pole structure is very
complicated. For $N=4$, a closed expression for \Npoint\ in terms
of $\Gamma$ functions is known \GSW:
\eqn\szero{A_{s=0}(k_1,..,k_4)=\pi\prod_{i=2}^4{\Gamma(k_1\cdot k_i-\beta_1
\beta_i+1)\over\Gamma(\beta_1\beta_i-k_1\cdot k_i)}}
It is easy to check that \szero\ satisfies factorization \resN. For
$N\geq5$ no such closed forms are known.

The situation is markedly different in two dimensions. There, the
singularity structure of \Npoint\ is much simpler, and the integral
representation can be evaluated for all $N$. The condition \mass\
implies that we have left and right moving massless ``tachyons''
in space-time:
\eqn\twodtach{T_k^{(\pm)}=\exp(ikX+\beta^{(\pm)}\phi);\;\beta^{(\pm)}=
-\sqrt{2}\pm k}
The correlation functions have then the following form
\ref\DFK{P. Di Francesco and D. Kutasov, Phys. Lett.
{\bf261B} (1991) 385.}:
\eqn\twodcor{\eqalign{
\langle T_{k_1}^{(+)}...T_{k_{N-1}}^{(+)}T_{k_N}^{(-)}\rangle
&=(N-3)!\prod_{i=1}^N(-\pi){\Gamma({1\over2}\beta_i^2-{1\over2}k_i^2)
\over\Gamma(1-{1\over2}\beta_i^2+{1\over2}k_i^2)}\cr
\langle T_{k_1}^{(+)}...T_{k_n}^{(+)}T_{p_1}^{(-)}...T_{p_m}^{(-)}
\rangle&=0;\;n,m\geq2\cr}
}
The $\Gamma$ factor for $i=N$ in \twodcor\ is infinite. This infinity
should be interpreted as the volume of the Liouville mode $\log\mu$
(see section 1). The vanishing of the second set of correlators in
\twodcor\ can be understood as a lack of such a volume factor. Taking
this into account, the first form properly understood describes all bulk
$N$ point functions \DFK, \DIIFK.

The results \twodcor\ are at first sight very puzzling. Both the
vanishing of amplitudes with $n,m\geq2$, and the simple form
of the amplitudes with $m=1$ (or $n=1$) are not at all obvious
from \Npoint; apriori one would expect {\it much more} poles, as in \resN.
We will not derive \twodcor\ here (the reader may find
a more complete discussion in \DFK, \DIIFK), but we will explain the
origin of its simple form. The phenomenon behind this simplicity
is partial decoupling of a certain infinite set of discrete states.
To understand why discrete states are important, we have to recall the
spectrum of the two dimensional string
\ref\SEI{N. Seiberg, unpublished.},
\PO,
\ref\conespec{B. Lian and G. Zuckerman, Yale preprint YCTP-P18-91 (1991);
S. Mukherji, S. Mukhi and A. Sen, Tata preprint TIFR/TH/91-25 (1991);
P. Bouwknegt, J. McCarthy and K. Pilch, preprint CERN-TH.6162/91 (1991).}.
We have already encountered
the tachyon field \twodtach. For $D>2$ there is in addition an infinite
tower of massive transverse oscillator states. For $D=2$ there are no
transverse directions, hence most of the massive degrees of freedom
are absent. There are still oscillator states at discrete values
of the momentum, which are related to oscillator primary states
of $c=1$ CFT
\ref\cVrs{J. Goldstone, unpublished; V. Kac, in {\it Group Theoretical
Methods in Physics}, Lecture Notes in Physics, vol. 94 (1979);
G. Segal, Comm. Math. Phys. {\bf 80} (1981) 301.}
(in space-time these are longitudinal oscillations of the
string). The general form of these states is:
\eqn\Vrs{V_{\rone,\rtwo}^{(\pm)}=\left[\partial^{\rone\rtwo}X+\cdots\right]
\exp\left(i{\rone-\rtwo\over\sqrt2}X+\beta_{\rone,\rtwo}^{(\pm)}\phi\right)}
where $r_1,r_2\in Z_+$, and the $\cdots$ stands for a certain polynomial
in $\partial X,\partial^2 X,\cdots$ (with scaling dimension $r_1r_2$).
The Liouville dressing is $\beta_{r_1,r_2}^{(\pm)}=-\sqrt{2}\pm{
r_1+r_2\over\sqrt{2}}$;
note that unlike \twodtach, here $V^{(+)}$ ($V^{(-)}$) correspond to
positive (negative) energy states.

The states \Vrs\ together with the tachyon comprise the full spectrum
of 2d string theory (at ghost number zero). Thus, all $N$ point functions
\Npoint\ must factorize on these states. To demonstrate the role of the
discrete states in \Npoint, we should analyze the factorization in various
channels. Consider, for example, the simplest
degeneration channel, when two tachyons collide. There are two possibilities
(up to $X\rightarrow -X$): 1) $T_{k_1}^{(+)}(z)T_{k_2}^{(-)}(0)$;
2) $T_{k_1}^{(+)}(z)T_{k_2}^{(+)}(0)$, $z\rightarrow0$.

In the first case, using the free OPE, which is clearly valid in bulk
correlators, although not in general \NATI, \JP, we find an infinite
set of poles at $k_1k_2+(k_1-\sqrt2)(k_2+\sqrt2)=-n$, $n=1,2,3,\cdots$.
The first pole ($n=1$) corresponds to an intermediate tachyon, while
the rest must have vanishing residues. The reason is that the residue involves
correlators of oscillator states (generalizing \resN),
but the momentum $k_1+k_2$ of the intermediate state is {\it not}
in the discrete list \Vrs\ (at least for generic $k_1, k_2$),
therefore the corresponding state must be BRST trivial (null), and decouple.

In the second case, the poles occur at $k_1k_2-(k_1-\sqrt{2})(k_2-\sqrt{2})=-n$
or: $\sqrt{2}(k_1+k_2)=2-n$. The intermediate momentum is automatically
in the discrete list \Vrs; furthermore $\beta=\beta_1+\beta_2=-\sqrt{2}
-{n\over\sqrt{2}}$ is such that the energy of the intermediate
state is negative -- it is a $V^{(-)}$. A similar situation occurs
for poles in more complicted channels.

Therefore, the pole structure
of \Npoint\ depends crucially on the physics of the states
$V_{\rone,\rtwo}^{(-)}$.
All but a finite number of the poles in the
tachyon S -- matrix correspond to
these states.
One can show \DIIFK\ that they decouple from
amplitudes of tachyons,
\eqn\vanish{\langle V_{\rone,\rtwo}^{(-)}T_{k_1}...T_{k_N}\rangle=0}
as long as $k_1,..., k_N\not\in {1\over\sqrt{2}}Z$. Therefore,
most of the poles in \Npoint\ have vanishing residues.
Using \vanish\ and factorization it is straightforward to show that most
of the tachyon amplitudes vanish, and the ones with signatures $(N-1,1)$
have the simple structure \twodcor\ (see \DIIFK).

The poles in \twodcor\ appear as ``external leg factors'' \PO, however
they also have a standard interpretation as poles in the $(i,N)$
channels (i.e. when $T_{k_i}$ and $T_{k_N}$ approach each other) --
the only channels where the poles have finite residues. Since $k_N$
is fixed by kinematics ($k_N=-{N-2\over\sqrt2}$), they appear to be
``leg poles''.
One could ask at this point
why these poles have a finite residue; after all, the intermediate
states are again $V^{(-)}$ in this case. The point is that the
decoupling \vanish\ of the discrete states with negative
energies is {\it only partial}. If there are enough $V^{(+)}$'s,
the amplitude does not vanish.
The residues of the poles in the $(i,N)$ channel involve the three
point function $\langle V_{\rone,\rtwo}^{(-)}T_{k_1}T_{k_2}\rangle$. By
the conservation laws, $k_1, k_2$ are in this case in the discrete list \Vrs\
and their energies are positive; this correlation function does not vanish.
The general situation is the following:
the most general correlator has the form:
\eqn\disc{\langle V_{\rone,\rtwo}^{(+)}...V_{r_{2n-1}, r_{2n}}^{(+)}
V_{s_1, s_2}^{(-)}... V_{s_{2l-1}, s_{2l}}^{(-)} T_{k_1}...T_{k_N}
\rangle\propto(\log\mu)^{n-l}}
where $k_1,...,k_N\not\in{1\over\sqrt{2}}Z$. Eq. \disc\
is very useful to determine the relative size of various
amplitudes. If the power of $\log\mu$ $(n-l)$ is one, we have a finite
bulk amplitude. This is the case for \tachcor\ with signature $(N-1,1)$
and generic $k_1,...,k_{N-1}$; $k_N$ is fixed by kinematics to
be $V_{0,N-2}^{(+)}$. If the power $n-l\leq0$, the bulk amplitude
(coefficient of $\log\mu$) vanishes. This is the case for tachyon
amplitudes with signature $(n,m)$, $n,m\geq2$, and generic momenta
(here the conservation laws do not constrain individual momenta),
and for correlation functions like \vanish.

One natural question that arises at this point is whether
we can reconstruct the general (non bulk) amplitudes in the theory from
the knowledge of the bulk ones. In general, this issue is not understood
(although it is probably possible to do this). However, in 2d (closed)
string theory it appears that one can make an educated guess for the answer,
based on a physical assumption. Since the results obtained by this
method are in agreement with those obtained from matrix models, the
physical assumption is probably justified,
however it has not been derived from first principles
(in the continuum formalism).
The idea is the following:
we saw before how the structure of 2d string theory leads to the
bulk correlators \twodcor. It is very natural to define ``renormalized
operators'' ($\Delta(x)\equiv{\Gamma(x)\over\Gamma(1-x)}$):
\eqn\redef{\tilde T_k={T_k\over(-\pi)\Delta({1\over2}\beta^2-{1\over2}k^2)}}
These have very simple correlation functions:
\eqn\simple{\langle\tilde T_{k_1}...\tilde T_{k_N}\rangle=(N-3)!}
What happens for non zero $s$ \Spar?
Integer $s>0$ is easy to treat since if $\langle\tilde T_{k_1}...T_{k_N}
\rangle=\tilde A_{k_1...k_N}\mu^s$, then differentiating $s$ times
w.r.t. $\mu$ we find
$$\tilde A(k_1,...,k_N)={\langle\tilde T_{k_1}...\tilde T_{k_N}(\tilde T_{
k=0})^s\rangle\over s!}$$
The numerator is a bulk amplitude, so we can use \simple\ and find:
$A(k_1, ...,k_N)={(s+N-3)!\over s!}\mu^s$, or:
\eqn\As{A_s(k_1,...,k_N)=({\partial\over\partial\mu})^{N-3}\mu^{s+N-3}}
We can further redefine the operators (and rescale the path integral) such
that all $\mu$ dependence on the r.h.s. of \As\ dissappears; alternatively
put $\mu=1$. Then we find ourselves in a situation where all correlators
at integer $s$ are polynomials in $k_1,...,k_N$ (using \Spar). The
question is how we can continue to non integer $s$. The first guess would
be to declare \As\ valid for any $s$ -- it certainly makes sense
for all amplitudes. While this is true in a certain region in momentum
space, in general the answer is more subtle.

There are several clues pointing in the direction we should choose.
The first clue is the fact that the effect of the massive modes
of the string is summarized, at least for bulk amplitudes, in the external
leg factors \twodcor. The renormalized field $\tilde T_k$ \redef\
is completely insensitive to the massive modes. This suggests that
the S -- matrix of $\tilde T$ is described by a local two dimensional
field theory.
One might have expected to see tachyon poles in amplitudes, however
as we saw, they are not there \twodcor. The reason is
lack of conservation of Liouville momentum in this field
theory \Spar. From the point of view of Lioville theory, we have the
OPE:
\eqn\LOPE{\exp(\beta_1\phi)\exp(\beta_2\phi)=\int d\beta\exp(\beta\phi)
f(\beta,\beta_1,\beta_2)}
where $f$ is an OPE coefficient. The contour of integration
for $\beta$ should \NATI\ run over macroscopic states $\beta=-{Q\over2}+ip$
($p\in R$). Using the OPE \LOPE\ in generic tachyon correlation functions
we find that the contribution of regions of moduli space
where vertices approach each other is:
\eqn\pint{\eqalign{
\int d^2z\langle T_{k_1}(z)T_{k_2}(0)...\rangle
\simeq&
\int_{\vert z\vert<\epsilon}d^2z\int_{-\infty}^\infty
dp(z\bar z)^{{1\over2}p^2+{1\over2}(k_1+k_2)^2-1}f(p, \beta_1,
\beta_2)\langle T_{k_1+k_2}...\rangle\cr
\simeq&
\int_{-\infty}^\infty dp{f(p,\beta_1,\beta_2)\over p^2+(k_1+k_2
)^2}\langle T_{k_1+k_2}...\rangle\cr}}
where we interchanged the order of the $z$ and $p$ integrations.
For fixed $p$, \pint\ has the familiar form from critical string
theory; we find a pole corresponding to the intermediate state
$T_{k_1+k_2}$. The fact that Liouville momentum is not conserved
and we have to sum over all $p$'s may turn this pole into a cut:
\pint\ depends on $\vert k_1+k_2\vert$. Thus we expect
cuts whenever some of the momenta $\{k_i\}$ in \tachcor\ satisfy
$\sum_ik_i\equiv p\rightarrow0$.
The fact that the bulk $\tilde T$ correlators are polynomial in $k_i$
\As\ is another indication that $\tilde T$ is described by a local
2d field theory. However, due to the expected appearance of cuts in amplitudes,
we have to be careful in continuing \As\ from its region of validity.
The remaining issue is to understand the expected analytic structure
of the correlation functions.

First, if the picture we have been developing is correct,
we expect the three point
functions $\langle T_{k_1}T_{k_2}T_{k_3}\rangle$ to be given
exactly by \As:
\eqn\threept{
\langle \tilde T_{k_1}\tilde T_{k_2}\tilde T_{k_3}\rangle=1}
The reason is that non analytic effects such as \pint\ can only occur
for $N\geq4$ point functions -- three point functions are insensitive
to the non-conservation of energy \LOPE.
This immediately allows us to obtain the propagator of the tachyon. By putting
$k_2=0$ in \threept\ and integrating we
obtain the two point function\foot{In critical string theory the two
point function vanishes, however, as explained above for the partition
sum, this is due to a division by the (infinite) volume of $\phi$. Here,
due to lack of translational invariance, the two point function
is finite, and  is related to the propagator.}:
\eqn\twopt{\langle \tilde T_k\tilde T_{-k}\rangle={1\over\sqrt{2}|k|}}
Hence the propagator (in a convenient normalization) should be:
${|k|\over\sqrt2}$. An important consistency check on this
is the comparison of an amplitude with an insertion
of a puncture $P=T_{k=0}$ to the amplitude without it. By
KPZ scaling \Spar\ we have:
\eqn\PTTT{\langle PT_{k_1}...T_{k_N}\rangle=\left[{1\over\sqrt2}
\sum_{i=1}^N\vert k_i\vert-(N-2)
\right]\langle T_{k_1}...T_{k_N}\rangle}
Thinking of \PTTT\ as a relation between tree amplitudes in the purported
space time field theory reveals its essential features: we can insert
the puncture $T_{k=0}$ into the tree amplitude $\langle T_{k_1}...T_{k_N}
\rangle$ either by attaching it to one of the $N$ external legs, thus adding
an internal propagator of momentum $k_i+0=k_i$ or inside the diagram.
The first term (the sum) on the r.h.s. of \PTTT\ corresponds to the first
possibility; we can read off the propagator
$|k|\over\sqrt2$,
which agrees with that obtained from \twopt.
The second term corresponds to the second possibility.
Now, after understanding the propagator, the remaining problem is specifying
the vertices in the space-time field theory. The three point vertex is 1
\threept. The higher vertices can be calculated using two basic
properties:

\noindent{}a) The form \As\ is exact for $k_1,...,k_{N-1}>0$, $k_N<0$,
which is the region in which all the energies involved are positive
\tachcor; from the world sheet point of view this is natural since
positive energy perturbations correspond to small deformations
of the surface \NATI, and also because the integrals \Npoint\ converge
there (after a certain analytic continuation in the central charge
\DIIFK).

\noindent{}b) All higher irreducible vertices are {\it analytic}
in $\{k_i\}$. To understand this, consider
for example the four point function:
\eqn\fourpt{\tilde A(k_1,..,k_4)=\int d^2z\langle\tilde T_{k_1}(0)
\tilde T_{k_2}(\infty)\tilde T_{k_3}(1)\tilde T_{k_4}(z)\rangle}
We can separate the $z$ integral
in \fourpt\ into two pieces. One is a sum of three contributions
from the regions $z\rightarrow0,1,\infty$.
By \pint\ we expect to get from those the tachyon
propagator\foot{The contribution of the massive states in the OPE
\pint\ is presumably related to the factors $\Delta({1\over2}
\beta_i^2-{1\over2}k_i^2)$ in \redef.}
${1\over\sqrt{2}}\vert k_1+k_i
\vert$. The rest of the $z$ integral is the contribution of the bulk
of moduli space (where $z$ is not close to $0,1,\infty$);
it gives a new irreducible four particle interaction
(which we will denote by $A_{1PI}^{(4)}$) for the
tachyons.
This contribution is analytic, since only massless intermediate states
cause cuts in the amplitudes \pint, and we have subtracted their
contribution.
Using this decomposition, we expect the following analytic structure
for $\tilde A(k_1,...,k_4)$:
\eqn\fourpnt{\tilde A({k_1,..,k_4})=
\left({\vert k_1+k_2\vert\over\sqrt{2}}
+{\vert k_1+k_3\vert\over\sqrt{2}}+{\vert k_1+k_4
\vert\over\sqrt{2}}\right)+A_{1PI}^{(4)}}
where $A_{1PI}^{(4)}$ is analytic in $k_i$. Next we use the fact that in the
region $k_1,k_2,k_3>0$, $k_4<0$, we actually know $\tilde A$ \As. Comparing
to \fourpnt\ we find
\eqn\afour{A_{1PI}^{(4)}=-1}
Eq. \afour\ holds now everywhere in $\{k_i\}$ by analytic continuation.

It is now clear how to proceed in the case of $N$ point functions. We assume
that we know already $A_{1PI}^{(4)}$,.., $A_{1PI}^{(N-1)}$.
Then we write
all possible tree graphs with $N$ external legs, propagator $
|k|\over\sqrt2$ and vertices
$A_{1PI}^{(n)}$
$(n\leq N-1)$ and add an unknown new irreducible
vertex $A^{(N)}_{1PI}
(k_1,..,k_N)$.
$A^{(N)}_{1PI}$ is again analytic in $\{k_i\}$
and we can fix it by comparing the sum of exchange amplitudes (reducible
graphs) and $A_{1PI}^{(N)}$
to the full answer \As\ in the appropriate
kinematic region. This fixes $A_{1PI}^{(N)}$
in the above kinematic
region. Now we use analyticity of
$A^{(N)}_{1PI}$ to fix it everywhere. The
outcome of this process is the determination
of the amplitudes in all kinematic regions given their values
in one kinematic region.

In practice, it is more convenient to obtain $A_{1PI}^{(N)}$
by Legendre transforming. This gives a highly non trivial set of
irreducible vertices; the general formulae are quite involved
\DIIFK. As an
example, one finds:
\eqn\threenz{ A_{1PI}^{(N)}(k_1,k_2,k_3,k_4=0,...,k_N=0)=
(\partial_{\mu})^{N-3}
\left\{{1\over\mu}
\prod_{i=1}^{3} { 1 \over \cosh( {k_i \over \sqrt{2}} \log \mu)}\right\}
\bigg\vert_{\mu=1} }
\eqn\fournz{
\eqalign{
A_{1PI}^{(N)}(k_1,k_2,k_3,k_4,0,...,0)&=\cr
(\partial_{\mu})^{N-4}
\mu^{-2} \left( \prod_{i=1}^4 {1 \over {\cosh({k_i \over \sqrt{2}}\log \mu)}}
\right)
&\left( -1 -\mu \partial_{\mu} \log \prod_{1 \leq i < j \leq 3}
\cosh({{k_i+k_j} \over \sqrt{2}} \log \mu) \right)\bigg
\vert_{\mu=1} \cr}}
As expected, the irreducible vertices are analytic in $\{k_i\}$. This is true
in general. Note that the preceeding discussion is very reminiscent
of the decomposition of moduli space which appears in general in closed
string field theory
\ref\zwie{H. Sonoda and B. Zwiebach, Nucl. Phys.
{\bf B331} (1990) 592; {\bf B336} (1990) 185.}. It would be
interesting to make the relation more precise. The tachyon field
theory constructed above is also closely related to the one which arises
in matrix models
\ref\SFT{S. Das and A. Jevicki, Mod. Phys. Lett. {\bf A5} (1990) 1639.},
\ref\SENG{A. Sengupta and S. Wadia,
Tata preprint TIFR/TH/90-13 (1990).},
\ref\GK{D. Gross and I. Klebanov, Nucl. Phys. {\bf B352} (1991) 671.}).

This concludes our presentation of tree level scattering in 2d closed
string theory. While we now know the structure of the S -- matrix in great
detail, our understanding of the origin of the factorization
of poles \twodcor\ and the emergence of the local tachyon field
theory is still unsatisfactory. For example, does the spectrum (massless
tachyon + discrete states) automatically lead to the structure we have
described? In particular, is the decoupling of the discrete states
with $E<0$ generic?

To gain more information about the possible structures in 2d string theory,
we have studied other string models with similar general characteristics.
The first is the 2d fermionic string. For $N=1$ supergravity \DIIFK,
the space-time dynamics includes two massless scalars
and a set of discrete states. The classical S -- matrix is extremely
similar to the bosonic case:  the discrete states with $E\leq0$
again (partially) decouple, and the bulk S -- matrix again
has only factorized poles, as in \twodcor.

The 2d $N=2$ (critical) string was studied in
\ref\Ntwo{H. Ooguri and C. Vafa, Mod. Phys. Lett. {\bf A5} (1990) 1389;
Nucl. Phys. {\bf B361} (1991) 469.},
and its tree level S -- matrix
is even simpler (all $N\geq 4$ point functions vanish). This is due
to the fact that the theory is really four dimensional.
Open 2d strings exhibit quite different behavior from the closed
case
\ref\DM{M. Bershadsky and D. Kutasov, Princeton preprint
PUPT-1283 (1991).}.
The spectrum contains again the massless field \twodtach\
and discrete states \Vrs, however the S -- matrix of
the open string tachyon $T_k^{(B)}$ on the disk has the form
($m_i={1\over2}\beta_i^2-{1\over2}k_i^2$):
\eqn\nm{\ao^{(n,m)}=\left[\prod_{i=1}^{n+m}{1\over\Gamma(1-m_i)}\right]
H_n(k_1,...,k_n)H_m(p_1,...,p_m)}
where
\eqn\hn{H_n(k_1,k_2,...,k_n)=
\sum_\sigma\prod_{l=1}^{n-1}{1\over\sin\pi\sum_{i=1}^l
m_{\sigma(i)}}}
and the sum over $\sigma$ in \hn\ runs over permutations of $1,2,...,n$.
The form \nm, \hn\ exhibits in general
a rich pole structure which is in essense due to the fact that the states
$V_{\rone,\rtwo}^{(-)}$ {\it do not} decouple here. Therefore, the decoupling
of the massive states from the tachyon dynamics also doesn't occur here.
Remarkably, the complicated S -- matrix \nm, \hn\ again follows from
a very simple space-time field theory. Indeed, one can check
\DM\ that the
generating functional:
\eqn\W{{\cal W}(T^{(\pm)})=\int D\psi D\psi^* D\bar\psi D\bar\psi^*
\exp\left[-\int dXd\phi e^{\sqrt{2}\phi}{\cal L}(\psi, T)\right]}
where the Lagrangian $\cal L$ is given by:
\eqn\Spsi{{\cal L}(\psi, T)=
\psi^*\sin({\pi\sqrt{2}}\partial^+)\psi+
\bar\psi^*\sin({\pi\sqrt{2}}\partial^-)\bar\psi
+T^{(+)}\bar\psi^*\bar\psi+T^{(-)}\psi^*\psi~,}
and $\partial^{\pm}=\partial_\phi \pm \partial_X$.
satisfies:
\eqn\gener{{\delta^{n+m}{\cal {W}}\over\delta T_{k_1}^{(+)}...
\delta T_{k_n}^{(+)}\delta T_{p_1}^{(-)}...\delta T_{p_m}^{(-)}}=
H_n(k_1,...,k_n)H_m(p_1,...,p_m)}
The reader should consult \DM\ for details and a
more precise formulation.
Hence, \W\ generates the scattering amplitudes \nm, \hn. Furthermore,
the propagator in \Spsi\ is essentially a lattice propagator, and it is
natural to define the theory on a space-time lattice in
${1\over2}(X\pm i\phi_l)$. Thus, the structure
arising in the open sector is both richer
\nm, \hn\ and simpler \W, \Spsi\ than the one observed in the closed sector
of the theory.

\bigskip

\newsec{COMMENTS.}

Despite the recent progress, there are still many unresolved fundamental
questions both in 2d string theory, and in its relation to higher
dimensional models. We have identified one feature which 2d string theory
shares with higher dimensional stable theories -- the fact that $c_{\rm
string}=1$ (the small effective number of states). It is not clear what other
features of 2d string theory have higher dimensional analogs. The non
critical superstring models of section 3 may be useful to study this issue.

Even within the framework of 2d string theory, the understanding
of the free fermion structure of \ORIG, \CONE\ is very incomplete. This
seems to boil down to a better understanding of the dynamics of
$V_{r,s}^{(\pm)}$ \Vrs. Understanding
the dynamics of the discrete
states is crucial also to study
gravitational physics in 2d string theory. The black hole of \WBH\
is described
\ref\BER{M. Bershadsky and D. Kutasov, Phys. Lett. {\bf 266B} (1991) 345.}
by turning on $V_{1,1}^{(-)}$; other
$V_{r,s}^{(-)}$ seem also to play a role in gravitational back reaction.

Two dimensional closed
string theory seems to be described by two consistent
(but different) S -- matrices. The S -- matrix for $T_k$ \tach, given
by \Npoint\ is sensitive to the discrete states \Vrs, and contains the
information on gravitational physics. Its pole structure and the effective
action which describes it may be used to study the space-time 2d gravity.
On the other hand, $\tilde T_k$ \redef\ are described by a second S -- matrix,
which originates from a two dimensional field theory for the tachyon field.
In particular, it doesn't contain space-time gravity. The appearance
of the local tachyon field theory is quite surprising from the point of
view of the continuum formulation (as is the case with many other
features of these models, it arises more naturally
in the matrix model \SFT, \SENG, \GK); it would
be interesting to understand this dual structure better.

A very important question concerns the form of the exact classical equations
of motion in 2d string theory. The correlation functions we have found
suggest a very interesting structure. In the sigma model approach, one
can investigate the moduli space of classical solutions by requiring
that adding perturbations to the world action:
${\cal S}\rightarrow {\cal S}+\lambda_\Delta\int V_\Delta$, does not spoil
conformal invariance
\ref\CFMP{C. Callan, D. Friedan, E. Martinec and M. Perry,
Nucl. Phys. {\bf B262} (1985) 593.}
\ref\BM{T. Banks and E. Martinec, Nucl. Phys. {\bf B294} (1987) 733.}.
{}From the form of the correlation functions of section 4 and matrix models
\MOORE, \INTER\ it follows that adding $\lambda_k\int T_k$ to the action
for all $k\not\in{1\over\sqrt{2}} Z$ and any $\lambda_k$ doesn't
spoil conformal invariance. We can calculate all correlators in a power
series in $\lambda_k$ (for fixed $\mu$), and find sensible
results. This seems to suggest that the exact non linear classical
equations of motion of 2d string theory have the property that a tachyon
field of the form $T(X)=\sum_{k\not\in{1\over\sqrt{2}}Z}T_k$ (and trivial
metric and other fields) is a solution. Of course, this can only be true
up to field redefinitions, but even then it is quite remarkable
(a special case of this is the claim that Liouville
\SL\
is a CFT).
When we turn on an expectation value of
a tachyon with one of the discrete momenta, the
metric and other fields back react. The details of this
back reaction haven't been
worked out yet.

The higher genus correlation functions are another issue, which
is still not resolved in the continuum formalism. In the matrix
model approach, all order correlation functions were obtained
in \MOORE. This is especially interesting in the light of the simple
results obtained; as an example, the two point function
was found to be given by:
\eqn\Gtwo{{\partial\over\partial\mu}\langle T_k T_{-k}\rangle=
{\Gamma(-\sqrt{2}|k|)\over\Gamma(\sqrt{2}|k|)}
{\rm Im}\left\{e^{{i\pi\over\sqrt{2}}|k|}\left[
{\Gamma({1\over2}+\sqrt{2}|k|-i\mu)\over\Gamma({1\over2}-i\mu)}
-{\Gamma({1\over2}-i\mu)\over\Gamma(-\sqrt{2}|k|+{1\over2}-i\mu)}\right]
\right\}}
Higher point functions appear in \MOORE. To extract the result
for given genus one should expand \Gtwo\ in powers of $1\over\mu^2$.
It is challenging to derive the higher genus correlation functions
in the continuum formalism, and even more challenging
to understand the origin of the ``non-perturbative'' results, such
as \Gtwo.

\bigbreak\bigskip\bigskip\centerline{{\bf Acknowledgements}}\nobreak

I would like to thank N. Seiberg for numerous discussions
of the issues presented here; part  of the work described
in this review
was done in collaboration with him, and the rest benefitted
greatly from his suggestions and criticism.
I would also like to thank M. Bershadsky and P. Di Francesco for
collaboration and discussions.
I have benefitted from discussions with
T. Banks, R. Dijkgraaf, J. Distler,
P. Freund, D. Gross, I. Klebanov,
E. Martinec, G. Moore, N. Seiberg, S. Shatashvili, S. Shenker
and H. Verlinde.
I am grateful to the organizers of the spring school and ICTP for their
hosptality.
This work was partially supported by
DOE grant DE-AC02-76ER-03072.

\listrefs
\end